\documentclass[aps, prd, 10pt, twocolumn, superscriptaddress,noshowpacs, preprintnumbers, longbibliography, bibnotes,hyperref,floatfix,nofootinbib]{revtex4-1}
\usepackage{subfigure}
\usepackage[colorlinks=true,breaklinks=true]{hyperref}
\hypersetup{allcolors=[rgb]{0.0 0.0 1.0},linkcolor=[rgb]{0.75 0.05 0.05}}
\usepackage[dvipsnames]{xcolor}
\usepackage{mathtools}
\usepackage{amsmath}
\usepackage{amsfonts}
\usepackage{amssymb}
\usepackage{bbold}
\usepackage{bm}
\usepackage{hyperref}
\usepackage{orcidlink}

\newcommand{\bD}{{\bf D}}

\newcommand{\bB}{{\bf B}}
\newcommand{\bL}{{\bf L}}
\newcommand{\bLam}{{\bf \Lambda}}
\newcommand{\bP}{{\bf P}}
\newcommand{\bR}{{\bf R}}
\newcommand{\br}{{\bf r}}
\newcommand{\bz}{{\bf z}}
\newcommand{\bS}{{\bf S}}
\newcommand{\bM}{{\bf M}}

\newcommand{\bJ}{{\bf J}}
\newcommand{\bG}{{\bf G}}
\newcommand{\vc}{v_{\rm c}}

\newcommand{\sD}{{\sf D}}

\newcommand{\GF}{G_{\rm F}}
\newcommand{\wP}{\omega_{\rm P}}

\long\def\exclude#1{}

\begin{document}

\title{Flavor solitons in dense neutrino gases}

\author{Damiano F.\ G.\ Fiorillo \orcidlink{0000-0003-4927-9850}} 
%\email{damianofg@gmail.com}
\affiliation{Niels Bohr International Academy, Niels Bohr Institute,
University of Copenhagen, 2100 Copenhagen, Denmark}

\author{Georg G.\ Raffelt
\orcidlink{0000-0002-0199-9560}}
%\email{raffelt@mpp.mpg.de}
\affiliation{Max-Planck-Institut f\"ur Physik (Werner-Heisenberg-Institut), F\"ohringer Ring 6, 80805 M\"unchen, Germany}

\begin{abstract}
We consider a dense neutrino gas in the ``fast-flavor limit''  (vanishing neutrino masses). For the first time, we identify exact solutions of the nonlinear wave equation in the form of solitons. They can propagate with both sub- or superluminal speed, the latter not violating causality. The soliton with infinite speed is a homogeneous solution and coincides with the usual fast-flavor pendulum except that it swings only once instead of being periodic. The subluminal soliton in the static limit corresponds to a one-swing ``spatial pendulum.'' A necessary condition for such solutions to exist is a ``crossed'' neutrino angle distribution. Based on the Nyquist criterion, we derive a new sufficient condition without solving the dispersion relation. The solitons are very fragile: they are as unstable as the homogeneous neutrino gas alone. Moreover, in the presence of matter, only the solution survives that is homogeneous in a frame comoving with the matter current. Generally, the matter effect cannot be eliminated by transformations in flavor space, but has real physical impact.
\end{abstract}

\date{21 March 2023}

\maketitle

\tableofcontents

\onecolumngrid
\vskip32pt
\twocolumngrid

\section{Introduction}

The daring idea to explain the missing solar neutrino flux by flavor evolution over astronomical distances \cite{Gribov:1968kq} was eventually correct, but initially  met fierce resistance from the belief that neutrinos should be massless. Today, determining the exact neutrino mass and mixing parameters has become a vast international effort of dedicated experiments and theoretical studies \cite{SajjadAthar:2021prg}. Ironically, recent theoretical studies of flavor evolution in neutrino-dense environments focus on massless neutrinos, the so-called fast-flavor limit \cite{Sawyer:2004ai, Sawyer:2005jk, Sawyer:2015dsa, Sawyer:2022ugt, Chakraborty:2016lct, Tamborra:2020cul, Volpe:2023met}. In this terminology, the usual mass-driven flavor conversion is called ``slow.'' It remains to be seen if fast-flavor conversion (FFC) indeed strongly impacts core-collapse supernovae or binary neutron-star mergers and concomitant nucleosynthesis
\cite{Tamborra:2017ubu, Abbar:2018shq, DelfanAzari:2019tez, Abbar:2019zoq, Glas:2019ijo, Nagakura:2019sig, Morinaga:2019wsv, Abbar:2020qpi, Capozzi:2020syn, Harada:2021ata, Kato:2021cjf, Zaizen:2021wwl, Nagakura:2021hyb, Richers:2021nbx, Richers:2021xtf, Abbar:2021lmm, Nagakura:2022kic, Capozzi:2022dtr, Richers:2022bkd, Grohs:2022fyq, Ehring:2023lcd, Nagakura:2023mhr}. Meanwhile, collective flavor waves in dense neutrino environments are a fascinating theoretical topic in their own right \cite{Raffelt:2011yb, Izaguirre:2016gsx, Capozzi:2017gqd, Abbar:2018beu, Capozzi:2018clo, Johns:2019izj, Martin:2019gxb, Capozzi:2019lso, DelfanAzari:2019epo, Johns:2020qsk, Bhattacharyya:2020jpj, Martin:2021xyl, Sigl:2021tmj, Morinaga:2021vmc, Johns:2021qby, Sasaki:2021zld, Dasgupta:2021gfs, Wu:2021uvt, Padilla-Gay:2021haz, Padilla-Gay:2022wck, Johns:2022yqy, Lin:2022dek, Xiong:2022vsy, Hansen:2022xza, Shalgar:2022rjj, Zaizen:2022cik,Kato:2022vsu, Bhattacharyya:2022eed, DedinNeto:2022xye, Xiong:2023upa}. New and surprising insights seem to emerge whenever one takes a fresh look.

One topic of considerable recent activity is the very question if the usual mean-field equations truly capture the flavor evolution of a dense neutrino gas, or conversely, if quantum entanglement of the many ``flavor spins'' is a dominant effect \cite{Rrapaj:2019pxz, Cervia:2019res, Roggero:2021asb, Roggero:2021fyo, Patwardhan:2021rej, Xiong:2021evk, Martin:2021bri, Cervia:2022pro, Roggero:2022hpy, Lacroix:2022krq, Siwach:2022xhx, Patwardhan:2023}. An early discussion~\cite{Bell:2003mg, Sawyer:2003ye} was apparently resolved in favor of the mean-field approach \cite{Friedland:2003dv, Friedland:2003eh, Friedland:2006ke}.  Whatever the final outcome of this debate, we here work on the purely refractive level of the mean-field approach, without heed for possible quantum limitations, but also without heed for the effect of collisions.

Arguably, FFC is the purest form of collective flavor evolution in that it does not require masses and mixing, and yet, in the mean-field approach, a dense neutrino gas supports a rich class of fast modes. In a recent paper \cite{Fiorillo:2023mze} we have shown a certain reciprocity between slow and fast modes so that much of our present discussion can be ported to the slow case as well. If fast-flavor waves actually occur in nature and how exactly they would be excited are actively studied questions. However, in line with our previous paper, we do not worry about phenomenological issues and continue to look at the subject from a mathematical perspective. 

Our main result is to identify a new class of solutions of the flavor wave equation in the form of \hbox{solitons}. These are exact nonlinear solutions that can exist with both sub- or superluminal speed.  Limiting cases are the traditional homogeneous flavor pendulum \cite{Johns:2019izj,Padilla-Gay:2021haz} and its static counterpart, a ``spatial pendulum'' \cite{Johns:2019izj}. Actually, the notion of a (temporal) soliton in this context was introduced in the parallel development concerning the Bardeen-Cooper-Schrieffer (BCS) Hamiltonian in condensed-matter physics \cite{Pehlivan:2011hp, yuzbashyan2005solution, Yuzbashyan:2005-PRB, yuzbashyan2005prb, yuzbashyan:2006, yuzbashyan2008normal, Yuzbashyan:2018gbu}. 

Another exact solution of the nonlinear flavor wave equation in the form of a uniformly moving flavor wave was recently discovered by Duan, Martin and Omanakuttan \cite{Duan:2021woc}. We believe that this solution corresponds to the homogeneous pure-precession solution found in Ref.~\cite{Raffelt:2007cb} as seen from a boosted frame in the same way as our superluminal soliton corresponds to the homogeneous flavor pendulum as seen from a boosted frame.

We assume that the reader is generally familiar with the subject of neutrino fast flavor oscillations and therefore limit our brief introduction to the essentials needed to establish the basic equations and  notation. 

The chosen environment is a homogeneous neutrino gas with an axisymmetric angle distribution. The coordinate along the symmetry direction is $r$ and the zenith-angle distribution is expressed through the velocity $v=\cos\theta$ of a given neutrino mode along $r$. In the fast-flavor limit, neutrino energy does not appear and the mean-field equations of motion (EOMs) are the same for neutrinos and antineutrinos. All relevant information is encoded in the lepton-number density matrix in flavor space ${\sD}_v(r,t)=\varrho_v(r,t)-\bar\varrho_v(r,t)$, where $\sD$ is a mnemonic for ``difference.'' We represent this matrix in the usual way by a Bloch vector $\bD_v(r,t)$ through ${\sD}_v-\frac{1}{2}\,{\rm Tr}\,\sD_v=\frac{1}{2}\bD_v\cdot{\bm\sigma}$ with $\bm\sigma$ a vector of Pauli matrices. After phase-space integration, we finally seek solutions of the EOM
\begin{equation}\label{eq:EOM-1}
    \left(\partial_t+v\partial_r\right)\,\bD_v=\mu
    \int_{-1}^{+1} dv'\,\left(1-vv'\right)\bD_{v'}\times\bD_v,
\end{equation}
where a dependence on $(r,t)$ is implied for $\bD_v$. The scale
$\mu=\sqrt{2}\GF(n_\nu+n_{\bar\nu})$ is a measure of the neutrino-neutrino refractive effect. The equivalent EOM in matrix form is
$i(\partial_t+v\partial_r)\,\sD_v=\mu \int  dv'\,(1-vv')[\sD_{v'},\sD_v]$.

This equation is more intuitive in linearized form where we assume that the off-diagonal elements of the density matrix (the $x$ and $y$ components of the Bloch vectors) are small. The $z$-component (the weak-interaction direction in flavor space) is represented by what has been called the ELN 
(for electron-lepton number carried by neutrinos) angle distribution or angular spectrum
\begin{equation}
    G_v=D_v^z=\int \frac{p^2 dp\,d\phi}{(2\pi)^3}\,
    \frac{(f_{\nu_e}-f_{\bar\nu_e})-(f_{\nu_\mu}-f_{\bar\nu_\mu})}{n_\nu+n_{\bar\nu}},
\end{equation}
where the occupation numbers depend on ${\bf p}$ expressed in polar coordinates through $p=|{\bf p}|$ and $v=\cos\theta$ and $\phi$. We have assumed that our two flavors are $e$ and $\mu$.

We denote the small off-diagonal element in the form
$\sD_v^{xy}=\frac{1}{2}\,(D_v^x-iD_v^y)=G_v{\cal D}_v$ so that the complex number
${\cal D}_v$ is a measure of flavor coherence and we assume $|{\cal D}_v|\ll1$. The linear EOM is consequently \cite{Izaguirre:2016gsx}
\begin{equation}
    i\left(\partial_t+v\partial_r\right){\cal D}_v=
   \mu\int_{-1}^{+1} dv'\,G_{v'}\left(1-vv'\right)\left({\cal D}_{v}-{\cal D}_{v'}\right).
\end{equation}
This EOM is similar to the Vlasov equation of plasma physics.

In the absence of neutrino-neutrino refraction ($\mu=0$), the Vlasov operator on the left-hand side simply causes a drift of any perturbation ${\cal D}_v$ (or wave packet) that may have been set up. The eigenfunctions are proportional to $\delta$ functions. For nonvanishing $\mu$, these so-called Case-Van Kampen modes \cite{VanKampen:1955wh} (or non-collective modes \cite{Capozzi:2019lso}) persist, although with modified singular eigenfunctions. Their dispersion relation is $\omega/k=v\leq1$, i.e., their phase velocity is subluminal, explaining their role in Landau damping of perturbations \cite{Sagan:1993es}.

In addition, new collective modes appear that do not exist without $\mu$ and, together with the non-collective modes, form a complete set. These collective modes can be stable or unstable, the latter being the modes that would engender FFC. Stable collective modes have a super-luminal phase velocity. It was Sawyer who first recognized that a nontrivial angle distribution $G_v$ can support unstable modes without neutrino masses \cite{Sawyer:2004ai, Sawyer:2005jk}. It was quickly understood that a ``crossing'' of $G_v$ is needed for an instability \cite{Izaguirre:2016gsx} and recently Morinaga proved that this is indeed a necessary and sufficient condition \cite{Morinaga:2021vmc}.

In the following we identify exact solutions of the nonlinear EOM that are connected to unstable solutions of the linear EOMs. In Sec.~\ref{sec:Temporal} we first recall the homogeneous pendulum solution and extend it to the nonperiodic limit of a soliton. As a next step, we transform it to a superluminally moving solution. In Sec.~\ref{sec:Spatial}, instead we consider a ``spatial pendulum'' and the nonperiodic limit of a static soliton. This in turn can be transformed to a subluminally moving solution. In Sec.~\ref{sec:Evolution} we turn to the stability question of these solutions. In Sec.~\ref{sec:MatterEffects} we show that in the presence of homogeneous matter, only the temporal soliton survives that is homogeneous in the Lorentz frame where the matter current is absent. A summary and conclusions are provided in Sec.~\ref{sec:Conclusion}. Some technical points are relegated to appendices. In particular, in Appendix~\ref{sec:SingleCrossed} we derive the conditions on $G_v$ required for a homogeneous instability (a pendulum solution) without solving the dispersion relation.

\section{Temporal Solitons}
\label{sec:Temporal}

Despite the non-linear nature of flavor evolution in a dense neutrino gas, under special conditions the evolution of the flavor composition may be surprisingly regular. In homogeneous settings, this behavior is equivalent to a pendular oscillation for single-crossed spectra~\cite{Johns:2019izj, Padilla-Gay:2021haz}; the regularity of the solution is here connected with the existence of non-trivial integrals of motion~\cite{Fiorillo:2023mze}. 
However, the introduction of inhomogeneous disturbances disrupts this regular behavior. Nevertheless, we demonstrate that there exist exact spatio-temporal solutions in which the flavor evolution is regular, taking the form of flavor solitons that propagate at a constant speed.

In this section, we first review the properties of the homogeneous evolution for single-crossed spectra and the analogy with pendular motion. We then show that the existence of homogeneous solutions implies, by Lorentz invariance, a special class of flavor solitons propagating with superluminal speed.

\subsection{Homogeneous Equations of Motion}

Our starting point is the EOM~\eqref{eq:EOM-1} in a homogeneous and axisymmetric neutrino gas. We write it in the more compact form
\begin{equation}\label{eq:EOM-2}
   \bigl(\partial_t +v\partial_r\bigr)\bD_v=(\bD_0-v\bD_1)\times\bD_v,
\end{equation}
where here and henceforth we absorb the interaction energy $\mu$ in the definition of dimensionless space and time coordinates. Moreover, following the previous literature \cite{Pehlivan:2011hp, Johns:2019izj, Fiorillo:2023mze, Raffelt:2007yz}, we define the moments of the distribution through
\begin{equation}\label{eq:moments-def}
   \bD_n=\sum_{v}\,v^n \bD_v.
\end{equation}
While this notation is slightly ambiguous because, for example, $\bD_0$ is the zeroth moment and not the mode $\bD_v$ with $v=0$, in practice there will be no confusion. 

Here and henceforth we usually consider a discrete set of velocities $\{v\}$ instead of a continuous distribution. This approach simplifies our discussion without changing the overall conclusions. The ``thermodynamic limit'' of a continuous distribution involves some subtleties that we have discussed in our earlier study~\cite{Fiorillo:2023mze}.

We assume that the system is initially homogeneous, including possible perturbations, and we are only seeking homogeneous solutions. With these assumptions, we first consider the EOMs
\begin{equation}\label{eq:homogeneous}
    \dot{\bD}_v=(\bD_0-v\bD_1)\times\bD_v.
\end{equation}
We assume the initial conditions are chosen with all $\bD_v$ nearly aligned to the flavor direction, taken as the $z$ axis. Summing on both sides over \smash{$\sum_v$} and \smash{$\sum_v v$} reveals that $\dot\bD_0=0$ and \hbox{$\dot\bD_1=(\bD_0+\bD_2)\times\bD_1$}. Therefore, the Bloch vector of total lepton number $\bD_0$ is conserved, whereas that of lepton-number flux $\bD_1$ follows a precession equation, implying that its length is conserved. 

\subsection{Pendulum Solutions}

In the unstable case, it is $\bD_1(t)$ that moves in analogy to a mechanical gyroscopic pendulum \cite{Johns:2019izj, Padilla-Gay:2021haz,Fiorillo:2023mze}. The functions $\bD_v(t)$ are then strongly correlated. Actually all of them are
linear superpositions of precisely three independent functions, which can be taken as $\bD_0(t)={}$constant, $\bD_1(t)$, and a third function $\bJ(t)$ that plays the role of the total angular momentum in the pendulum analogy. Alternatively, the three vector functions can be taken as three ``carrier modes'' \cite{Raffelt:2011yb} or ``auxiliary spins'' \cite{yuzbashyan2005solution}, which are special cases of the ``Lax vectors'' of this system as we have recently explained \cite{Fiorillo:2023mze}. 

In polar coordinates, the motion of $\bD_1(t)$ is fully represented by its zenith angle $\vartheta(t)$ and azimuth angle $\varphi(t)$ in the $x$-$y$-plane.\footnote{For polar coordinates in the space of Bloch vectors, we use $\vartheta$ and $\varphi$, whereas in coordinate space, we use $\theta$ and $\phi$ so that $v=\cos\theta$.} So eventually the dynamics of the full system $\bD_v(t)$ is encapsulated in two scalar functions $\vartheta(t)$ and $\varphi(t)$, which in turn are universal and depend on only two parameters: the natural frequency and spin of the equivalent pendulum. The explicit analytic expressions for these functions are derived in Appendix~\ref{sec:gyropendulum}.

For a given initial configuration $\bD_v$ aligned with the $z$-direction and with the spectrum
$G_v=D_v^z(0)$, the first step toward a pendulum solution is stability analysis. In a normal-mode expansion, one seeks solutions that initially behave as $e^{-i\Omega t}$. The eigenfrequency $\Omega$
must obey the dispersion relation obtained by linearization \cite{Izaguirre:2016gsx,Padilla-Gay:2021haz}
\begin{equation}
    \sum_v \frac{v^2 G_v}{v G_1-G_0+\Omega}=1,
\end{equation}
where the $n$-th moments of the spectrum are
\begin{equation}
    G_n=\sum_v G_v v^n.
\end{equation}    
Recall that $\bD_0$ and $|\bD_1|$ are conserved. An alternative form is
\begin{equation}\label{eq:dispersion_time}
    \sum_v \frac{v G_v}{v G_1-G_0+\Omega}=0,
\end{equation}
which is equivalent to the presence of a Lax vector with vanishing $z$-component \cite{Fiorillo:2023mze}.

An instability requires a spectral crossing, i.e., $G_v$ must change sign for some $-1<\vc<+1$. At present, we are only examining spectra that have a single crossing. This necessary condition would also be sufficient if we were to include inhomogeneous normal modes, i.e., those with a non-vanishing wave number \cite{Morinaga:2021vmc}. For the homogeneous case, a sufficient condition based on the Nyquist criterion is (Appendix~\ref{sec:SingleCrossed})
\begin{equation}\label{eq:Nyquist-lab}
    \frac{G_1}{G_0\vc}<0
    \quad\hbox{and}\quad
    \int dv\,\frac{v G_v}{G_0(v-\vc)}<0.
\end{equation}
This criterion is directly related to the Penrose criterion for the instability of a single-humped electron distribution in a collisionless plasma~\cite{penrose1960electrostatic} (see, e.g., Ref.~\cite{sturrock1994plasma}); the mathematical connection between collisionless plasma and fast flavor conversions in the linear regime was already noted in Ref.~\cite{Fiorillo:2023mze}. Therefore, one can assess the presence of an instability without solving the dispersion relation. 

A system $\{\bD_v,v\}$ consisting of a set of Bloch vectors on a discrete set of velocities can be degenerate in that for given initial conditions, the solution fills a smaller-dimensional phase space. For our standard case of all $\bD_v$ initially aligned and a single instability, one can construct a system with the same $\bD_1(t)$ consisting of any number of beams $N\geq 3$ that could be smaller or larger than the original system. Starting from a supernova-inspired continuous (or finely binned quasi-continuous) spectrum $G_v$, we can construct an infinity of three-beam representations to achieve the same $\bD_1(t)$, or directly a pendulum representation. This enormous degeneracy is attributed
to the large number of invariants as explained, e.g., in our previous paper~\cite{Fiorillo:2023mze}. In Appendix~\ref{sec:flavor_pendulum} we show explicitly how to construct a three-beam representation from a given $G_v$ and conversely, how to construct a larger-dimensional system starting from a three-beam one or directly from a chosen pendulum.

\begin{table}[b!]
 \caption{Three-mode reference cases.}
    \vskip4pt
    \label{tab:examples}
    \centering
    \begin{tabular*}{\columnwidth}{@{\extracolsep{\fill}}lllllll}
    \hline\hline
    Case &\multicolumn{3}{l}{Spectrum} & Moments      &Instability        &Pendulum\\
         & $v_1$     & $v_2$     & $v_3$     & $G_0$ & $\omega_{\rm P}$  & $\lambda$ \\
         & $G_{v_1}$ & $G_{v_2}$ & $G_{v_3}$ & $G_1$ & $\Gamma$          & $\sigma$ \\
     \hline
     A   & $-1$      & $+0.5$    & $+1$ & $-0.4$ & $-0.5$ & $+0.6$     \\
         & $-1$      & $-0.4$    & $+1$ & $+1.8$ & $+0.332$ & $-0.833$     \\  
     B   & $-1$      & $+0.5$    & $+1$ & $-0.9$ & $-0.563$ & $+0.9$     \\
         & $-1.125$  & $-0.9$    & $+1.125$ & $+1.8$ & $+0.703$ & $-0.625$   \\
     \hline
    \end{tabular*}
    \vskip12pt
\end{table}

As a matter of convenience, we will frequently use a few explicit three-beam examples with properties defined in Table~\ref{tab:examples}. For three beams with velocities $v_i$ and lengths $G_{v_i}$, the dispersion relation is quadratic in $\Omega$ and thus easily solved. It yields the real and imaginary components of the eigenfrequency $\Omega-D_0=\wP+i\Gamma$ 
\begin{subequations}\label{eq:three-mode-frequency}
    \begin{eqnarray}
        &&\wP   = \frac{J}{2},
        \\
        &&\Gamma = \frac{1}{2}\sqrt{4G_0G_1v_1v_2v_3-J^2},
    \end{eqnarray}
\end{subequations}
where we introduce the total angular momentum of the pendulum $J=G_2-G_1 \sum_i v_i$. Notice that initially, when all $\bD_v$ are aligned, $J=S$, the pendulum spin. 
Following Eq.~\eqref{eq:pendulumpars-App}, the natural pendulum frequency and spin parameter are 
\begin{equation}\label{eq:pendulumpars}
    \lambda = \sqrt{\omega_{\rm P}^2+\Gamma^2}
    \quad\hbox{and}\quad
    \sigma = \frac{\omega_{\rm P}}{\sqrt{\omega_{\rm P}^2+\Gamma^2}}.
\end{equation}
In terms of the three-beam parameters, they are
\begin{subequations}\label{eq:three-mode-pendulum-pars}
    \begin{eqnarray}
        \lambda   &=& \sqrt{G_0 G_1 v_1 v_2 v_3}
        \\
%        \sigma&=& \frac{J}{2\sqrt{G_0 G_1 v_1 v_2 v_3}}
\sigma&=& \frac{G_2-G_1(v_1+v_2+v_3)}{2\sqrt{G_0 G_1 v_1 v_2 v_3}}
    \end{eqnarray}
\end{subequations}
We show the numerical values for our examples in Table~\ref{tab:examples}.

\begin{figure}[htb!]
    \centering
    \includegraphics[width=0.80\columnwidth]{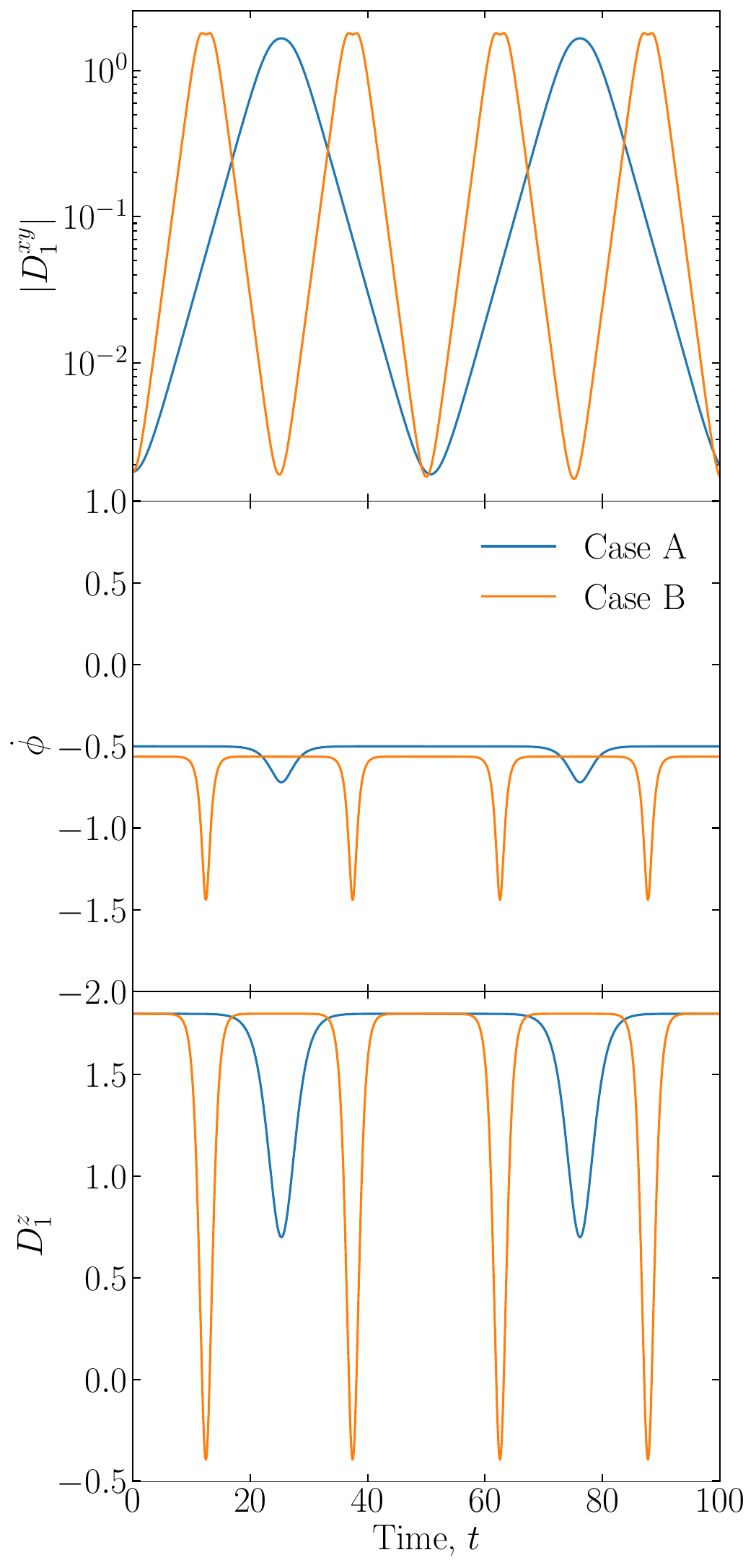}
    \caption{Components of $\bD_1(t)$ for our two benchmark cases of three-beam systems defined in Table~\ref{tab:examples}.}
    \label{fig:periodicsoliton}
%\end{figure}
\vskip20pt
%\begin{figure}[htb]
    \centering
    \hbox to\columnwidth{\includegraphics[height=0.48\columnwidth]{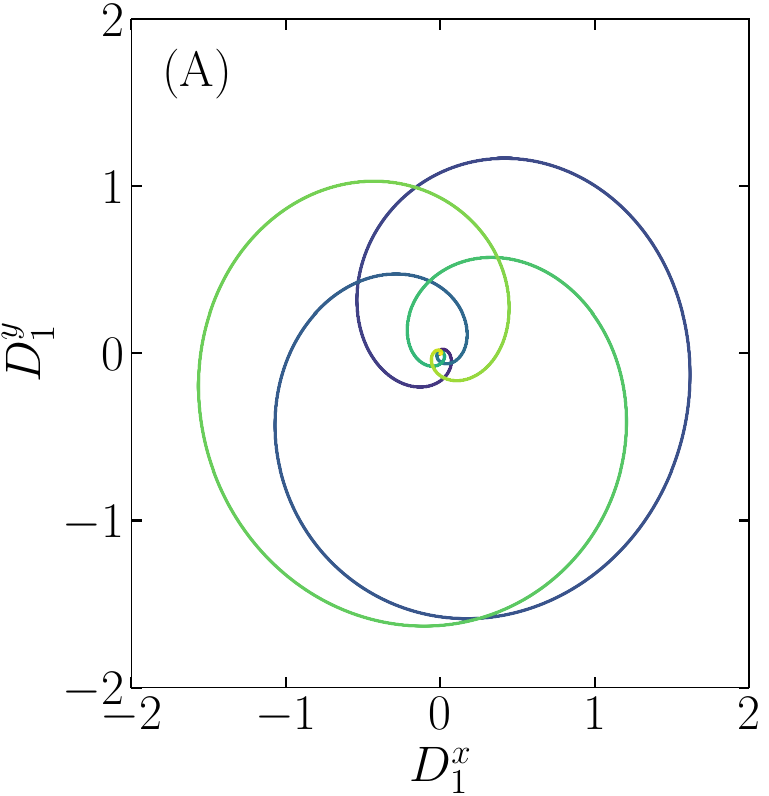}\hfill
    \includegraphics[height=0.48\columnwidth]{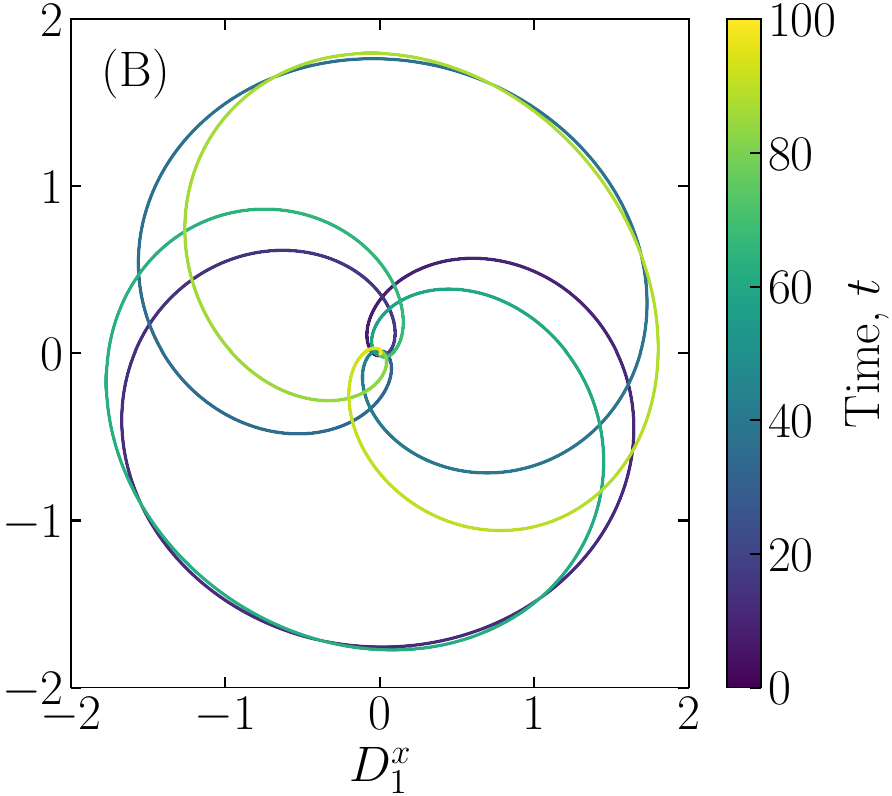}}
    \caption{Pendulum trajectories in the $D^x_1$--$D^y_1$ plane for the two benchmark cases defined in Table~\ref{tab:examples}. Shown are two zenith-angle periods. The azimuth-angle period is generally not commensurate, causing the trajectory to fill the plane after many zenith periods.}
    \label{fig:periodicsolitontrajectory}
\end{figure}

Figure~\ref{fig:periodicsoliton} shows the trajectories for these examples in terms of the vertical component $D^z_1$; the in-plane component $|D^{xy}_1|=\sqrt{(D^x_1)^2+(D^y_1)^2}$; and the azimuth precession velocity $\dot{\varphi}$, expressing $\bD_1$ in polar coordinates as $\bD_1=D_1(\sin\vartheta\cos\varphi,\sin\vartheta\sin\varphi,\cos\vartheta)$. If initially all $\bD_v$ were perfectly aligned, the system would be stuck in an unstable fixed point, so we have provided a small disturbance as a seed. The component $D^z_1$ shows a periodic behavior, describing the periodic swing of the pendulum from the upright position back to it. Likewise, the transverse component (the modulus of the $x$-$y$-component) is periodic, where the long waiting periods of $D_1^z$ are accompanied by exponential growth of $D_1^{xy}$, or exponential decline on the approach back to the upward position. 

On the other hand, the pendulum trajectory is \textit{not} periodic in general, as one can see in Fig.~\ref{fig:periodicsolitontrajectory}, where we show the projection of the pendulum trajectory in the \hbox{$D^x_1$--$D^y_1$} plane for two periods of $D_1^z$. The motion is actually conditionally periodic: $\vartheta(t)$ and $\varphi(t)$ are
separately periodic, however with non-commensurate periods unless the seed was fine-tuned.
We still refer to the pendular motion as periodic, with the implication that it is only periodic in the zenith and azimuth angle separately. The azimuth motion $\varphi(t)$ does not vanish even in the upright position except for vanishing spin (plane pendulum). The motion $\varphi(t)$ is 
never retrograde (Appendix~\ref{sec:gyropendulum}), the overall direction being determined by the initial spin orientation.

\subsection{Homogeneous Temporal Soliton}

\begin{figure}[htb!]
    \centering
\includegraphics[width=0.80\columnwidth]{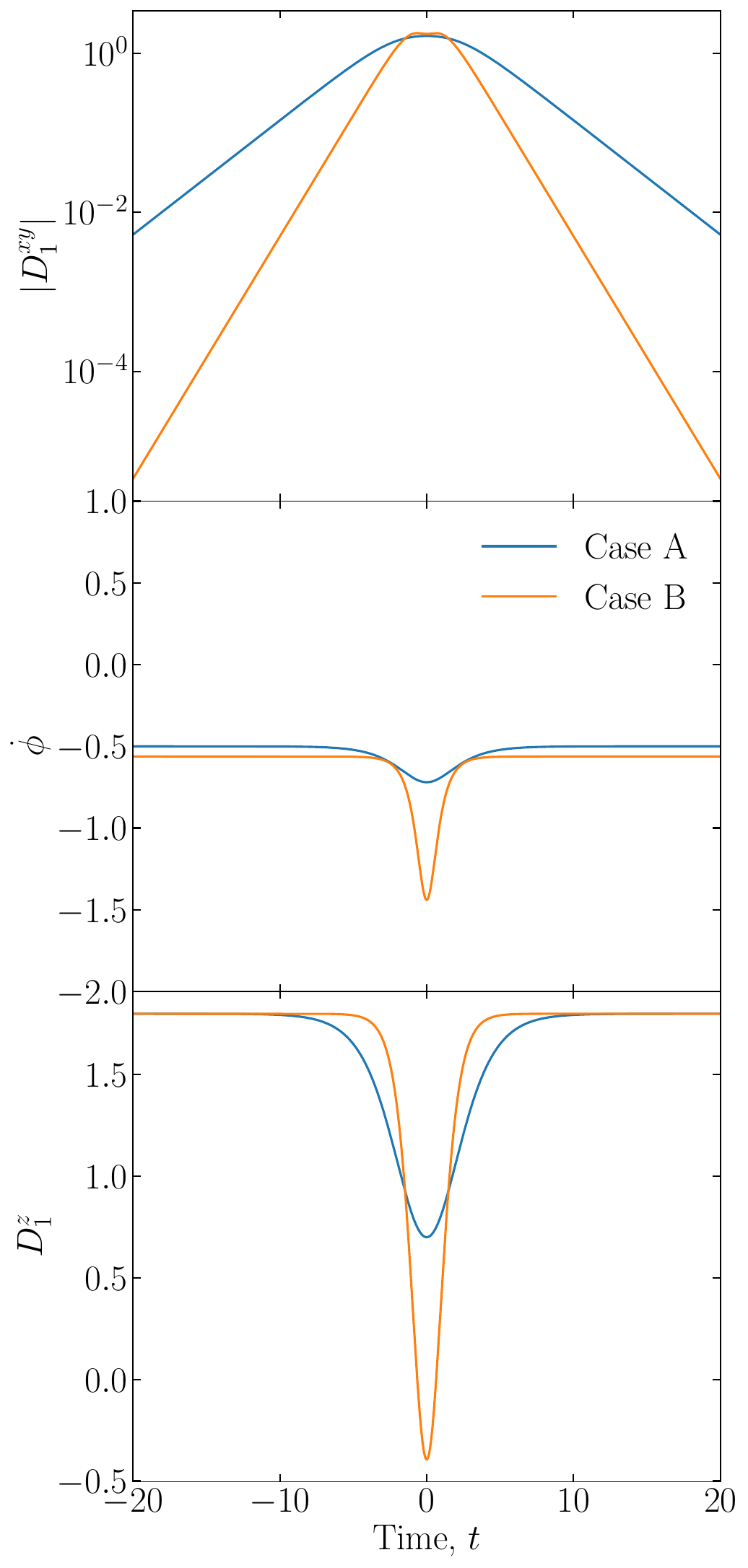}
    \caption{Components of $\bD_1(t)$ for a temporal soliton for our benchmark cases with parameters given in Table~\ref{tab:examples}. The $D_1^z$ component stays flat for $t\to\pm\infty$, whereas the transverse components continue to shrink exponentially for $t\to\pm\infty$ towards the unstable fixed point.}
    \label{fig:singlesoliton}
%\end{figure}
\vskip20pt
%\begin{figure}[htb]
    \centering
    \hbox to\columnwidth{\includegraphics[height=0.48\columnwidth]{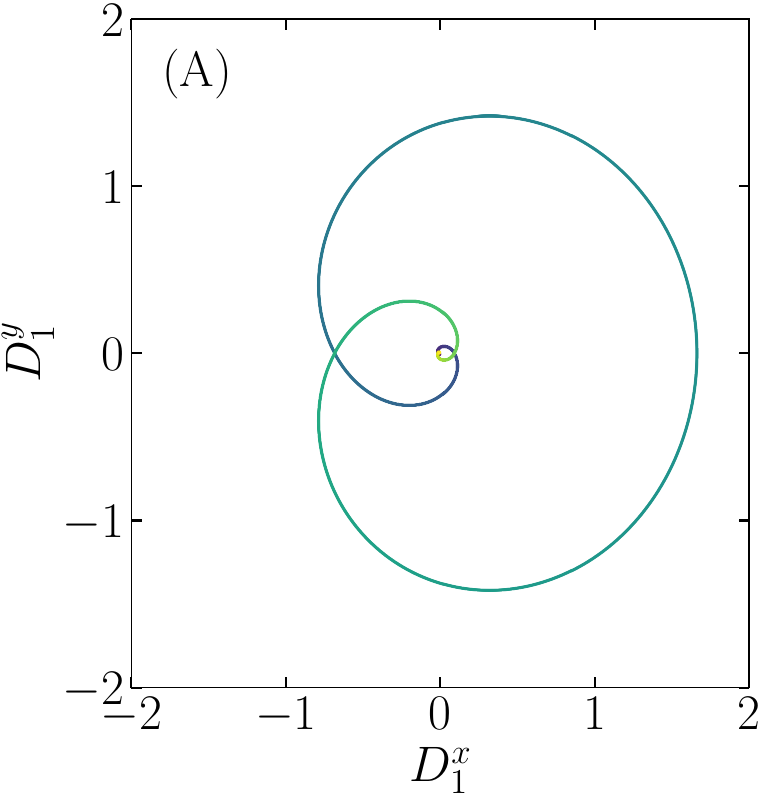}\hfill
    \includegraphics[height=0.48\columnwidth]{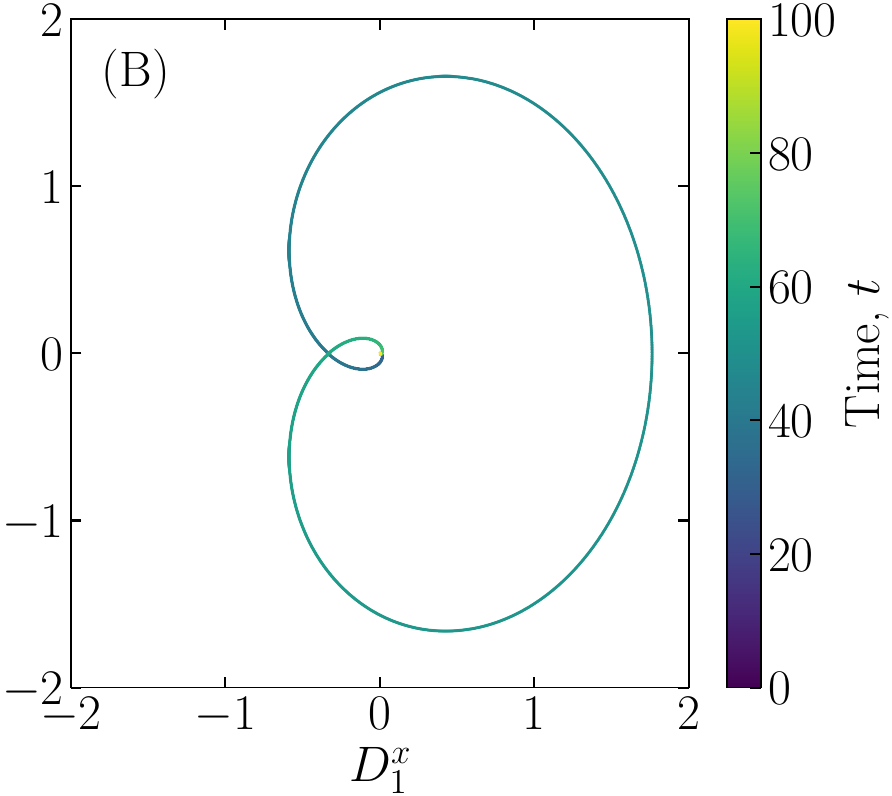}}
    \caption{Soliton trajectories in the $D^x_1$--$D^y_1$ plane for the two benchmark cases defined in Table~\ref{tab:examples}.}
    \label{fig:singlesolitontrajectory}
\end{figure}

The full motion of the pendulum then consists of a periodic swing of the zenith angle. The separation between two full swings, the waiting period, is determined by the amplitude of the initial small perturbation. If the latter is sufficiently small, we may regard the swings as nearly separate motions. We can then construct an exact solution in which the perturbation is infinitely small at infinitely remote times, such that the motion is effectively a single pendulum swing. 

To acquaint the reader with our later terminology, we call this motion a ``temporal soliton,'' although this is not a soliton in the usual sense -- it does not involve any spatial variation. Our definition matches the one used in Ref.~\cite{yuzbashyan2008normal} to describe the time evolution of the order parameter in a BCS superconductor, a system which we have shown to follow the same equation as the fast-flavor conversions in a homogeneous setting~\cite{Fiorillo:2023mze}. 

A more fitting notion might be that of an ``instanton'': a singular event on the infinite time axis. However, we avoid this terminology because its quantum-field theory connotations could be misleading. The temporal soliton, a pendulum that swings only once, approaches its classical unstable fixed point at $t\to\pm\infty$, without ever reaching it, of course.

We construct the soliton analytically in Appendix~\ref{sec:gyropendulum}, based on the pendulum equations of motion, and fixing time $t=0$ when the zenith angle is 
at its lowest point. While this approach avoids the need for a seed (an initial perturbation), instead one needs to fix the time when this event happens. Besides the choice of this instant, the pendulum motion $\vartheta(t)$ and $\varphi(t)$ is fixed by only two parameters, the natural pendulum frequency $\lambda$ (in units of $\mu$, the neutrino-neutrino interaction energy, that we usually absorb in the definition of time) and the spin expressed as $S=2\lambda\sigma$. The spin parameter 
$\sigma$ is defined such that $0\leq\sigma^2\leq1$ for an instability to exist. For larger $\sigma^2$, the pendulum is stuck in the ``sleeping top'' upright position.

Expressing $\bD_1$ in terms of polar coordinates as discussed earlier, and using $c(t)=\cos\vartheta(t)$, the soliton is (Appendix~\ref{sec:gyropendulum})
\begin{eqnarray}\label{eq:soliton}
    \varphi(t)&=&\sigma\lambda t
    +\arctan\biggl[\frac{\sqrt{1-\sigma^2}}{\sigma}\,\tanh\left(\sqrt{1-\sigma^2}\,\lambda t\right)\biggr],
    \nonumber\\[1ex]
    c(t)&=&-1+2\Bigl[\sigma^2+(1-\sigma^2)\tanh^2\left(\sqrt{1-\sigma^2}\,\lambda t\right)\Bigr].
    \nonumber\\
\end{eqnarray}
We may regard the soliton as the elementary object which is periodically repeated in the solutions of Figs.~\ref{fig:periodicsoliton} and~\ref{fig:periodicsolitontrajectory}. For a soliton, we show in Figs.~\ref{fig:singlesoliton} and~\ref{fig:singlesolitontrajectory} the components $D^z_1$, $|D^{xy}_1|$, and $\dot{\varphi}$, as well as the trajectories projected in the $D^x_1$--$D^y_1$ plane.

In summary, for a spectrum $G_v$ with an instability in the linearized system, there exists a temporal soliton. Except for the instant when it happens, its properties are all fixed by the linear eigenfrequency $\Omega=\wP+i\Gamma$.

\subsection{Multiple Solitons}

The above discussion relates only to angular distributions with a single unstable mode. If more than one unstable mode exists, that can happen for a multi-crossed spectrum, each instability defines a different soliton, that can happen at an arbitrary instant. If they happen at very different times, they are essentially two different solitons. If they happen more closely to each other, the common solution includes complicated interference effects in the time period of overlap. Explicit examples have been worked out in the BCS context in  Ref.~\cite{yuzbashyan2008normal}.

Such multiple soliton solutions, based on multiple instabilities, are to be distinguished from the periodic pendulum motion. The latter is not a sequence of events but rather a single periodic solution. 

\subsection{Uniformly Moving Soliton}
\label{sec:superluminal}

An exact solution that depends both on space and time can now be found by identifying solutions that are homogeneous in a boosted frame, which we identify as primed. These solutions therefore obey $\partial\bD_v/\partial r'=0$. We call $V$ the speed of the frame in which the solution is homogeneous, and $\gamma=(1-V^2)^{-1/2}$ is the corresponding Lorentz factor. Notice that the original neutrino gas, having a non-isotropic angle distribution, does not define a natural laboratory frame. In any moving frame, it is once more a homogeneous and non-isotropic gas, but with a transformed single-crossed spectrum $G_v$. 

However, instead of transforming $G_v$ explicitly, the EOM can be directly mapped to the EOM of the original system, where the soliton was homogeneous. The simplest way to do so is to use the covariant form of the EOMs
\begin{equation}\label{eq:covariant_eom}
    k^\mu \partial_\mu \bD_v=\bD^\mu\times\bD_v k_\mu,
\end{equation}
with $k^0=1$ and $k^1=v$ in the laboratory frame and $\bD^\mu=\sum_{v}\bD_{v}k^\mu$. In the primed (moving) frame, $k^0=\omega'=\gamma(1-v V)$ and $k^1=k'=\gamma(v-V)$. Since $\partial\bD_v/\partial r'=0$, the EOMs in the primed frame read
\begin{equation}
    \frac{\partial \bD_v}{\partial t'}=\left(\bD'_0-\frac{k'}{\omega'}\bD'_1\right)\times \bD_v,
\end{equation}
after dividing everywhere by $\omega'$.

We further redefine $v'=k'/\omega'=(v-V)/(1-vV)$ as the velocity in the new frame. The range $-1\leq v\leq1$ of course maps on $-1\leq v'\leq1$ as behooves a Lorentz transformation, and conversely
$v=(v'+V)/(1+v'V)$. Moreover, we redefine the Bloch vectors in the new frame  
\begin{equation}\label{eq:redef_bloch}
\bS_{v'}=\bigl[\omega'\bD_v\bigr]_{v'=k'/\omega'}
    =\bigl[\gamma(1-vV)\bD_v\bigr]_{v=\frac{v'+V}{1+v'V}}
\end{equation}
so as to match the definitions of the moments $\bD'_n=\sum_{\xi'} \xi'^n \bS_{\xi'}$. Thus, we finally reach the form
\begin{equation}
    \frac{\partial \bS_{v'}}{\partial t'}=(\bD'_0-v'\bD'_1)\times\bS_{v'},
\end{equation}
which is identical to Eq.~\eqref{eq:homogeneous}.

We can now perform a linear stability analysis around the asymptotic state with polarization vectors closely aligned to the $z$-axis. Since the EOMs are identical to Eq.~\eqref{eq:homogeneous}, we may simply perform the appropriate replacements in Eq.~\eqref{eq:dispersion_time}. Assuming a solution behaving as $e^{-i\Omega' t'}$, we obtain
\begin{equation}\label{eq:dispersion_superluminal}
    \sum_v \frac{(1-vV)(v-V)G_v}{vG_1-G_0+\gamma(1-vV)\Omega'}=0,
\end{equation}
where $G_v$ is the original spectrum and $G_0$ and $G_1$ its moments in the original frame. Whenever this dispersion relation admits a complex solution for some $|V|<1$, the solution in the corresponding reference frame will behave as the homogeneous swing of the pendulum, and, in the limit of infinitely small initial perturbation, approaches the temporal soliton. In this comoving frame, the vector $\bD'_1(t')$ in all space performs a single pendulum swing.

In the laboratory frame, the solution depends then only on the combination $t'=\gamma(t-V r)$, or, equivalently on the combination $r-t/V$. We recognize this as a uniformly moving wavefront with the velocity $v_\mathrm{soliton}=1/V$. Since $|V|<1$, it follows that the corresponding soliton is superluminal. The homogeneous temporal soliton is obtained from here in the limit $V=0$, which formally corresponds to a soliton moving with infinite speed. 

\begin{figure}[b!]
    \centering
    \hbox to\columnwidth{\includegraphics[height=0.84\columnwidth]{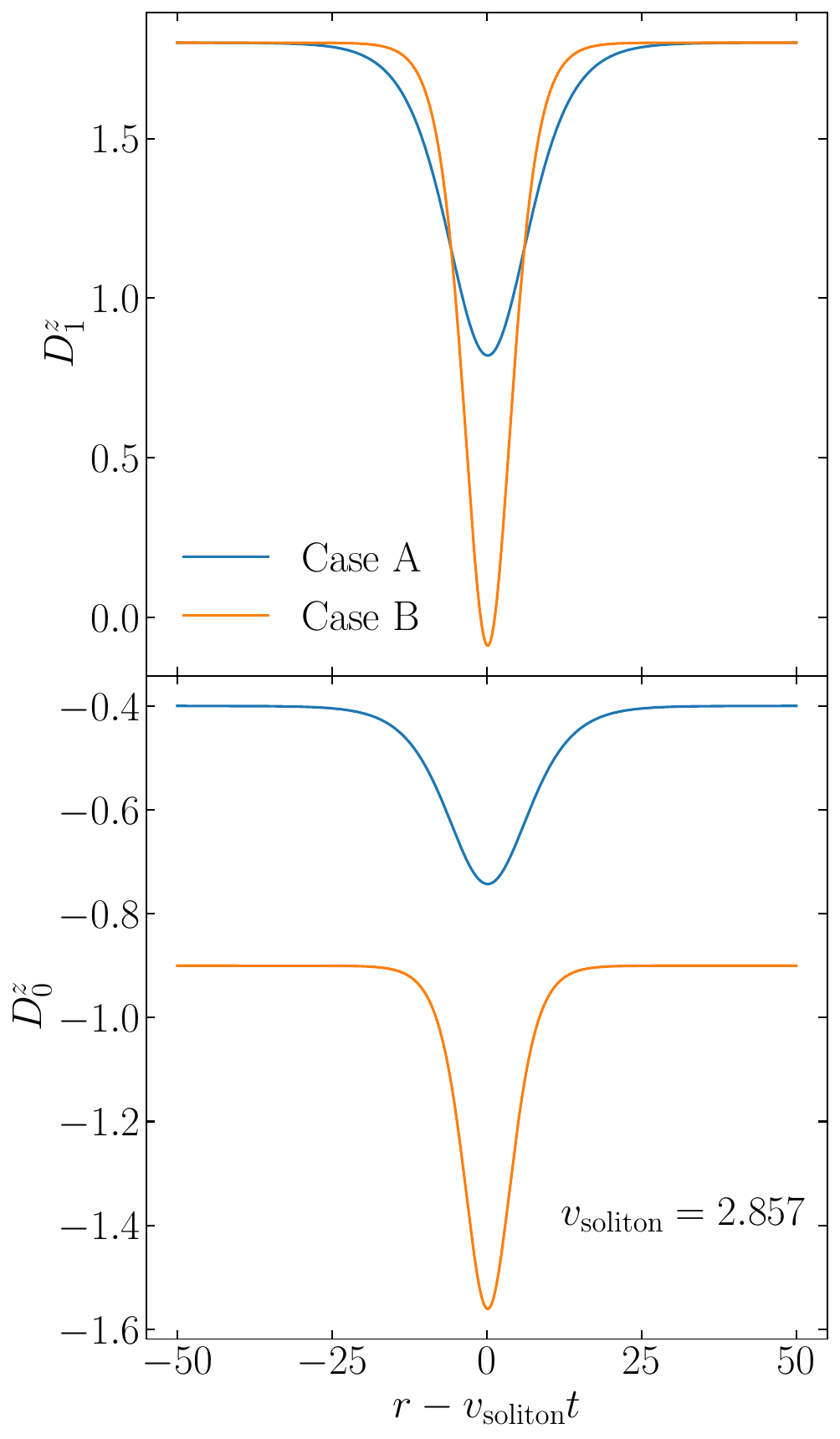}
    \hfil
    \includegraphics[height=0.84\columnwidth]{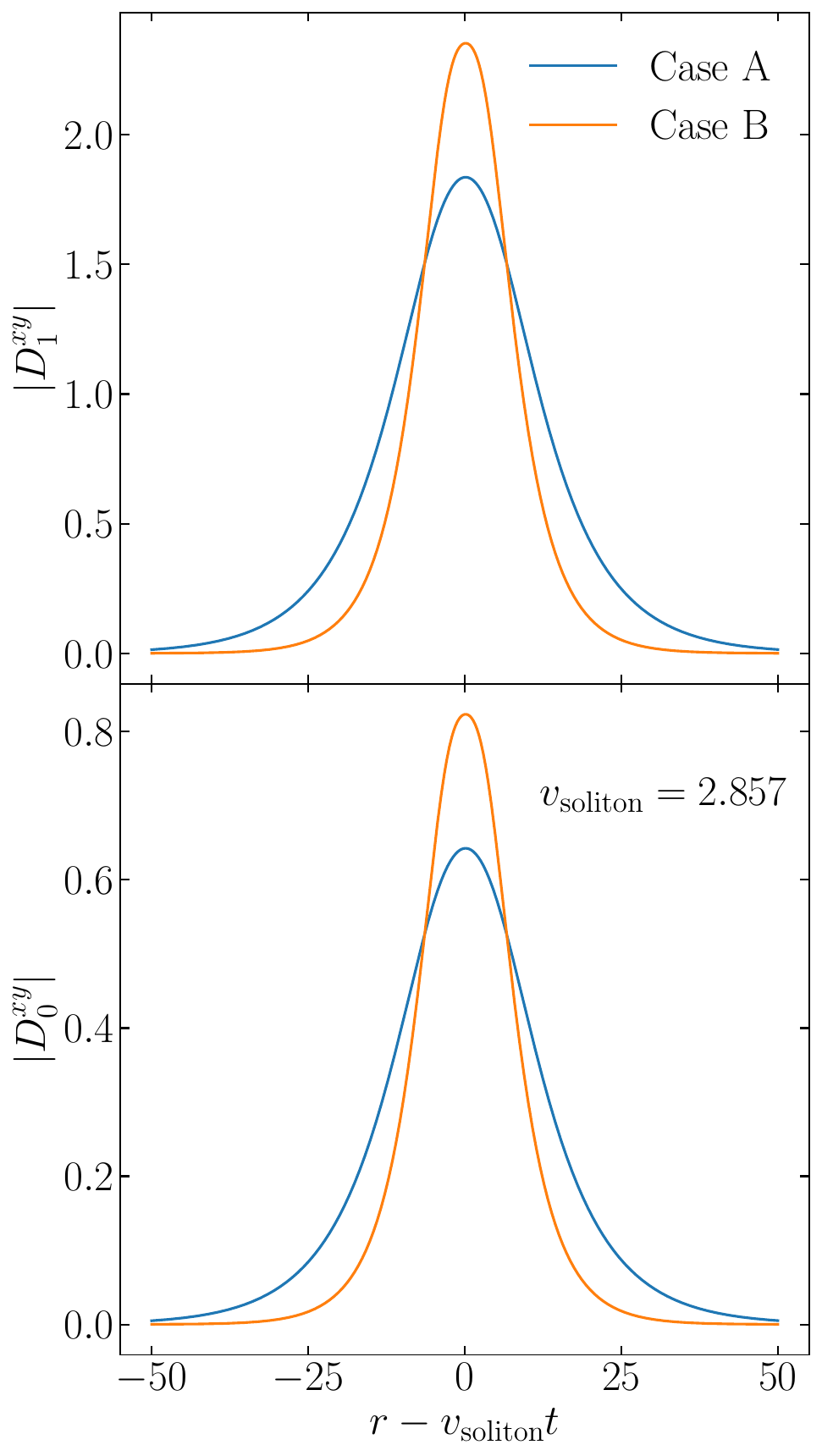}}
    \caption{Structure of the superluminal soliton as a function of the comoving coordinate $r-v_\mathrm{soliton} t$ for a superluminal soliton velocity $v_\mathrm{soliton}=V^{-1}=2.857$, for the two reference cases defined in Table~\ref{tab:examples}. Here $V=0.35$ is the speed of the frame in which the soliton is homogeneous. In the original frame, $\bD_0$ and $|\bD_1|$ are no longer conserved.}
    \label{fig:superluminalsolitonz}
\end{figure}

Figure~\ref{fig:superluminalsolitonz} shows the spatial structure of $\bD_1$ and $\bD_0$ for the superluminal soliton for both cases A and B defined in Table~\ref{tab:examples} for the example $V=0.35$, corresponding to $v_{\rm soliton}=V^{-1}=2.857$. Notice that, while $\bD'_0$ and the pendulum length $|\bD'_1|$ are spatially and temporally constant, the lab-frame moments $\bD_0=\gamma(\bD'_0+V\bD'_1)$ and $\bD_1=\gamma(\bD'_1+V\bD'_0)$ depend both on space and time, and their length is not constant.

At each instant, the flavor composition is close to the asymptotic one far from the soliton, while it develops off-diagonal coherence in the soliton region which moves with superluminal speed $v_\mathrm{soliton}$. The spatial width of the soliton is directly connected with the duration of a swing of the flavor pendulum in the comoving frame in which it is homogeneous. Assuming this duration to be of the order of $T'$, the lab-frame spatial width is of the order of $\Delta=T'/\gamma V=T' \sqrt{v_\mathrm{soliton}^2-1}$, and therefore becomes arbitrarily large as $v_\mathrm{soliton}\to\infty$. This of course corresponds to our previous statement that the limit of $v_\mathrm{soliton}\to\infty$ leads to the homogeneous, temporal soliton.

\subsection{Violation of Causality?}

The superluminal motion raises of course questions on the validity of this solution, in view of causality requirements. However, causality is broken already by the initial conditions which spawn this solution. This is most easily seen in the limit $V=0$, which corresponds to the homogeneous temporal soliton: the homogeneity of the initial flavor configuration over arbitrarily large distances requires a correlation which cannot have been set up by a causal physical process. 

In the same way, the superluminal soliton corresponding to a non-zero $V$ does not involve any superluminal propagation of information. Indeed, the information on the soliton existence at a certain position $r$ does not originate from the motion of the soliton at previous times, but is rather already contained in the initial conditions in the past light cone of the point $r$. Therefore, if we chose an initial condition at $t=0$ which is identical to the soliton for $r>0$ and with an arbitrary shape for $r<0$, at a later time $t$ we would still see the exact soliton solution for any $r>t$, even though the soliton itself was valid only over half the space at the initial time. 

This situation is analogous to what happens in dielectric materials with a superluminal group velocity, where an initial wavepacket can propagate with a superluminal group velocity. This is examined in detail, e.g., in Ref.~\cite{Diener:1996mj}, where it is shown explicitly that this superluminal propagation does not correspond to a real transmission of information. 

\subsection{Dispersion Relation}

A necessary condition for a neutrino gas to support a soliton is for the spectrum $G_v$ to be crossed, and this property is invariant against Lorentz boosts along the symmetry axis. On the other hand, the sufficient condition 
stated in Eq.~\eqref{eq:Nyquist-lab} must be evaluated in the new frame and may not have a solution. In other words, not every chosen $V$ provides a soliton, it may rather provide a stable ``sleeping top'' configuration. The boosted conditions for an instability are provided in Appendix~\ref{sec:SingleCrossed} for the case of a continuous spectrum $G_v$.

For our discrete three-mode examples, one may simply evaluate the dispersion relation of Eq.~\eqref{eq:dispersion_superluminal} directly. In analogy to the lab-frame case, it is a quadratic equation for $\Omega'$. For our example~A, one finds a non-vanishing imaginary part for $-0.099 < V< 0.40$, for
B the range is $-0.43 < V< 0.43$. Even though the two extreme neutrino beams have $v=\pm1$, the allowed range for $V$ does not exhaust this full range.

\begin{figure}[t!]
    \centering
    \includegraphics[width=0.75\columnwidth]{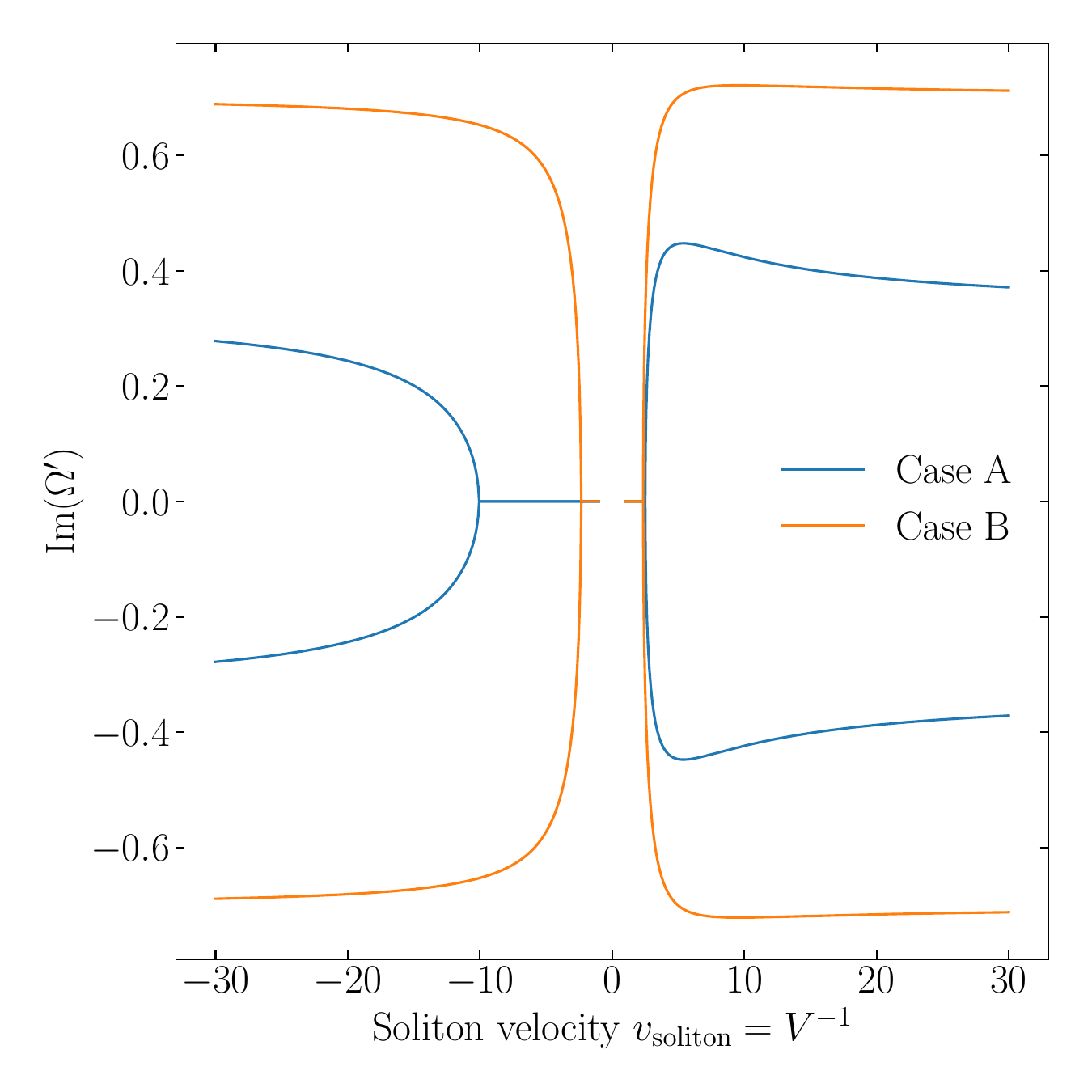}
    \vskip-6pt
    \caption{Imaginary part of the comoving-frame frequency $\Omega'$ as a function of the soliton speed $v_\mathrm{soliton}=V^{-1}$ for the two benchmark cases defined in Table~\ref{tab:examples}.    }
    \label{fig:dispersion_superluminal}
\end{figure}

Conversely, only a certain range of speeds $v_\mathrm{soliton}=V^{-1}$ is supported. We display Im$\,\Omega'$ as a function of $v_\mathrm{soliton}$
for both of our reference cases in Fig.~\ref{fig:dispersion_superluminal}. The soliton is supported only for sufficiently large speeds, corresponding to the range in which Im$(\Omega')\neq0$.

Finally, the eigenfrequency of the linear system corresponding to the soliton is $\Omega=\gamma\Omega'$ in the laboratory frame, and the corresponding
wavevector is $K=\gamma\Omega'V$, implying
\begin{equation}
    \frac{\Omega}{K}=\frac{1}{V}
    \quad\text{with}\quad
    -1<\frac{1}{V}<1.
\end{equation}
Therefore, the superluminal soliton corresponds to linear eigenmodes with complex $\Omega$ {\it and}
complex $K$ such that
\begin{equation}\label{eq:RealRatioTemporal}
    \text{Im}\left(\frac{\Omega}{K}\right)=0
    \quad\hbox{and}\quad
    \frac{\Omega}{K}>1
    \quad\hbox{or}\quad
     \frac{\Omega}{K}<-1.
\end{equation}
In other words, starting from the inhomogeneous dispersion relation
\cite{Izaguirre:2016gsx}, one needs to find solutions
$(\Omega,K)$ that have the same complex phase and their ratio corresponds to a super-luminal phase velocity. 

This is different from the usual normal-mode analysis where one looks for a complex $\Omega$, assuming a perturbation with a real wave number. Instead, we are here looking for a complex $\Omega$ as a function of the 
velocity~$V$, a real parameter.

\section{Spatial solitons}

\label{sec:Spatial}

The original flavor pendulum (of slow oscillations), like so many other neutrino oscillation phenomena, was actually meant to represent a spatial phenomenon, an evolution along a direction in space \cite{Duan:2005cp, Hannestad:2006nj, Duan:2007mv}. On the other hand, our earlier fast temporal instability and concomitant soliton was a purely temporal event until it emerged, under a boost, as a superluminal form of nonlinear wave propagation. Conversely,
one can contemplate a purely spatial fast-flavor pendulum, an instability along a direction in space \cite{Johns:2019izj}. We here show that in similar manner such a ``spatial instability'' can be seen as a static flavor soliton, now using this terminology in a more proper sense. Later we continue with the now-familiar boost that turns up yet new nonlinear solutions, the class of subluminally moving solitons.

\subsection{Static Soliton}

To get started, we search for exact solutions of the static EOMs
\begin{equation}\label{eq:eom_static}
    v\frac{\partial\bD_v}{\partial r} =(\bD_0-v\bD_1)\times\bD_v.
\end{equation}
By dividing everywhere by $v$, this becomes 
\begin{equation}
    \frac{\partial\bD_v}{\partial r} =(v^{-1}\bD_0-\bD_1)\times\bD_v.
\end{equation}
We now introduce the modified definitions
\begin{equation}
    \bM_v=\bD_v v
    \quad\text{and}\quad
    \bM_n=\sum_v \bM_v v^{-n},
\end{equation}
so that $\bM_0=\bD_1$ and $\bM_1=\bD_0$. The EOMs then take the form
\begin{equation}
    \frac{\partial\bM_v}{\partial r}=(v^{-1}\bM_1-\bM_0)\times\bM_v,
\end{equation}
which, except for a change in sign, has the same form as Eq.~\eqref{eq:homogeneous}, an observation first made by Johns et al.~\cite{Johns:2019izj}. Compared to the temporal case, in which the variable $v$ runs in the compact interval from $-1$ to $1$, in this EOM the role of the variable $v$ is played by $v^{-1}$, which lies in the non-compact interval $|v^{-1}|>1$. However, this modification does not lead to significant differences in the physics and the nature of the solutions. 

Assuming that asymptotically the polarization vectors are all aligned with the $z$-axis, just as in the temporal case we can determine the instabilities of the system using the linear dispersion relation. Assuming solutions behaving in space as $e^{iKr}$, we can deduce the dispersion relation directly by the analogy with Eq.~\eqref{eq:dispersion_time}. After the appropriate substitutions are made, we find 
\begin{equation}\label{eq:dispersion_spatial}
    \sum_v \frac{vG_v}{G_0-v G_1+Kv}=0.
\end{equation}
As we show in Appendix~\ref{sec:SingleCrossed}, if a system admits an unstable frequency from Eq.~\eqref{eq:dispersion_time}, it also admits an unstable wavevector in Eq.~\eqref{eq:dispersion_spatial}.

Linear stability analysis in the spatial case deserves some clarification. In the temporal case, causality automatically determines that frequencies with a positive imaginary part lead to a temporal instability. In the spatial case, causality does not provide such a guide, and solutions with a complex $K$ may lead both to a growth of the solution or to a damping. Which of these conditions is realized from a given boundary conditions set at an initial time depends on the dispersion relation at $\Omega\neq0$~\cite{pitaevskii2012physical}. However, here we are not interested in the topic of whether growth can arise from a given boundary condition. Rather, we simply show the existence of exact soliton solutions of Eq.~\eqref{eq:eom_static} if 
the dispersion relation provides a complex solution for $K$. Therefore, we do not pursue the topic of boundary conditions for a neutrino gas any further.

Since Eq.~\eqref{eq:eom_static} has been mapped to the same form as Eq.~\eqref{eq:homogeneous}, we conclude that the solutions we found in the previous sections are also solutions of Eq.~\eqref{eq:eom_static}, with a change of interpretation. The temporal solution, with a single swing of the pendulum $\bD_1$, corresponds here to a static, localized region in which the ``spatial'' flavor pendulum $\bM_1=\bD_0$ swings away from the $z$ axis. We dub this a spatial soliton and corresponds to the intuitive picture of some sort of localized wave packet.
In this region, the neutrino flavor density matrix has large off-diagonal coherence, whereas outside, it returns to the asymptotic flavor composition. Notice that here it is $\bM_0=\bD_1$ that remains constant throughout space. 

We stress that the solution is static even though individual neutrinos are still moving with their own velocities $v$ in the direction of the symmetry axis. Neutrinos passing through the soliton region develop off-diagonal coherence under the influence of the mean field of all the other neutrinos in the same region, and return to their original flavor structure after they pass through the soliton.

In the temporal case, a given initial condition generally leads to a periodic swing of the polar angle of the pendulum, as we discussed above. In the spatial case, the analogous structure is a periodic lattice of solitons; we remind the reader that we use periodicity here with the caveat that only the polar angle is periodic, while the azimuthal angle periodicity is not in general commensurate with the polar angle one.

The soliton existence relies on a delicate balance between the advection of neutrinos and their non-linear interaction. Furthermore, as we have seen, a diagnostic of the static soliton existence is the presence of a complex $K$ solution to the dispersion relation. Indeed, the static soliton is the non-linear evolution of the eigenmodes with $\Omega=0$ and complex $K$ in the dispersion relation, in the same sense that the temporal soliton is the non-linear evolution of the eigenmodes with complex $\Omega$ and $K=0$. 

\subsection{Uniformly Moving Soliton}

In similar manner to the temporal case, we can identify a new class of solutions which correspond to the static soliton in a boosted frame. Our procedure mirrors the one to obtain the superluminal solitons in Sec.~\ref{sec:superluminal}. We start from the covariant form of the EOMs, Eq.~\eqref{eq:covariant_eom}, and write them explicitly in the frame in which the solution is static
\begin{equation}
    v'\frac{\partial \bD_v}{\partial r'}=\left( \bD'_0-v' \bD'_1 \right)   \times\bD_v.
\end{equation}
As we did in Sec.~\ref{sec:superluminal}, we redefine the Bloch vectors in the new frame with Eq.~\eqref{eq:redef_bloch}, obtaining
\begin{equation}
v'\frac{\partial \bS_{v'}}{\partial r'}=(\bD'_0-v'\bD'_1)\times\bS_{v'}.
\end{equation}
This is now identical to Eq.~\eqref{eq:eom_static}, with the variable $v'$ still defined in the range between $-1$ and $1$. Therefore, it admits solutions which behave as a static soliton in the boosted reference frame. Such solutions depend only on the variable $r'=\gamma(r-Vt)$. Therefore, in the laboratory frame, they are solutions depending both on space and time, corresponding to a soliton moving with the subluminal velocity $V$.

Linearization around the asymptotic state with all vectors closely aligned with the $z$ axis provides the dispersion relation for $K'$, the wavevector in the soliton rest frame, assuming a solution $\propto e^{i K' r'}$. After replacing the correct mapping in Eq.~\eqref{eq:dispersion_time} we find
\begin{equation}
    \sum_v\frac{(1-vV)(v-V)G_v}{D^z_0-vD^z_1+\gamma(v-V)K'}=0.
\end{equation}
Whenever this equation admits complex solutions for some value of $V$, then a uniformly moving soliton exists as a solution of the general EOMs. Notice that the wavenumber as seen from the laboratory frame is $K=\gamma K'$. Furthermore, the frequency in the laboratory frame is $\Omega=\gamma K' V$. Therefore, in analogy to
Eq.~\eqref{eq:RealRatioTemporal} 
we may characterize the moving solitons as the nonlinear evolution of excitations with complex $K$ and
\begin{equation}
    \text{Im}\left(\frac{\Omega}{K}\right)=0
    \quad\text{and}\quad
    -1<\frac{\Omega}{K}<1.
\end{equation}
So once more, these are solutions of the dispersion relation with equal complex phase for $\Omega$ and $K$, this time with subluminal phase velocity that here plays the role of the soliton speed.

\section{Soliton stability}

\label{sec:Evolution}

The soliton solutions are exact, in the sense that if the initial conditions are precisely set, their evolution follows a uniform motion in the laboratory frame. A natural question is then whether they are also stable, namely if they will survive tiny deviations from the exact initial conditions.

The instability of the single-soliton solution can be grasped by noting that the asymptotic state of the soliton is already by itself unstable. For a subluminal soliton, we can always choose a frame in which the solution is a static soliton; in this frame, the asymptotic state is certainly unstable against homogeneous perturbations, as we show in Appendix~\ref{sec:SingleCrossed}. Thus, the soliton carries within itself the seed of its own destruction, since its own existence requires an instability of its own asymptotic state.

A full stability analysis of the soliton is, however, difficult to carry out analytically, since the unperturbed state, consisting in a single soliton, is not translationally invariant. Therefore, even in the linearized regime, elementary perturbations cannot be looked for in the form of plane waves, making impossible a spatial Fourier analysis in this context. For this reason, we limit ourselves to a numerical study of the soliton stability.

Our procedure is therefore to solve the full spatio-temporal EOMs~\eqref{eq:EOM-2}, assuming as an initial condition the structure of a soliton moving with velocity $V=0.1$, evaluated at time $t=0$. Since the initial condition is set on a numerical grid, the discreteness of the grid acts as a natural source of perturbation for the subsequent evolution, and therefore we do not insert any additional seed. 

\begin{figure}[b!]
\vskip-6pt
    \centering
    \includegraphics[width=0.85\columnwidth]{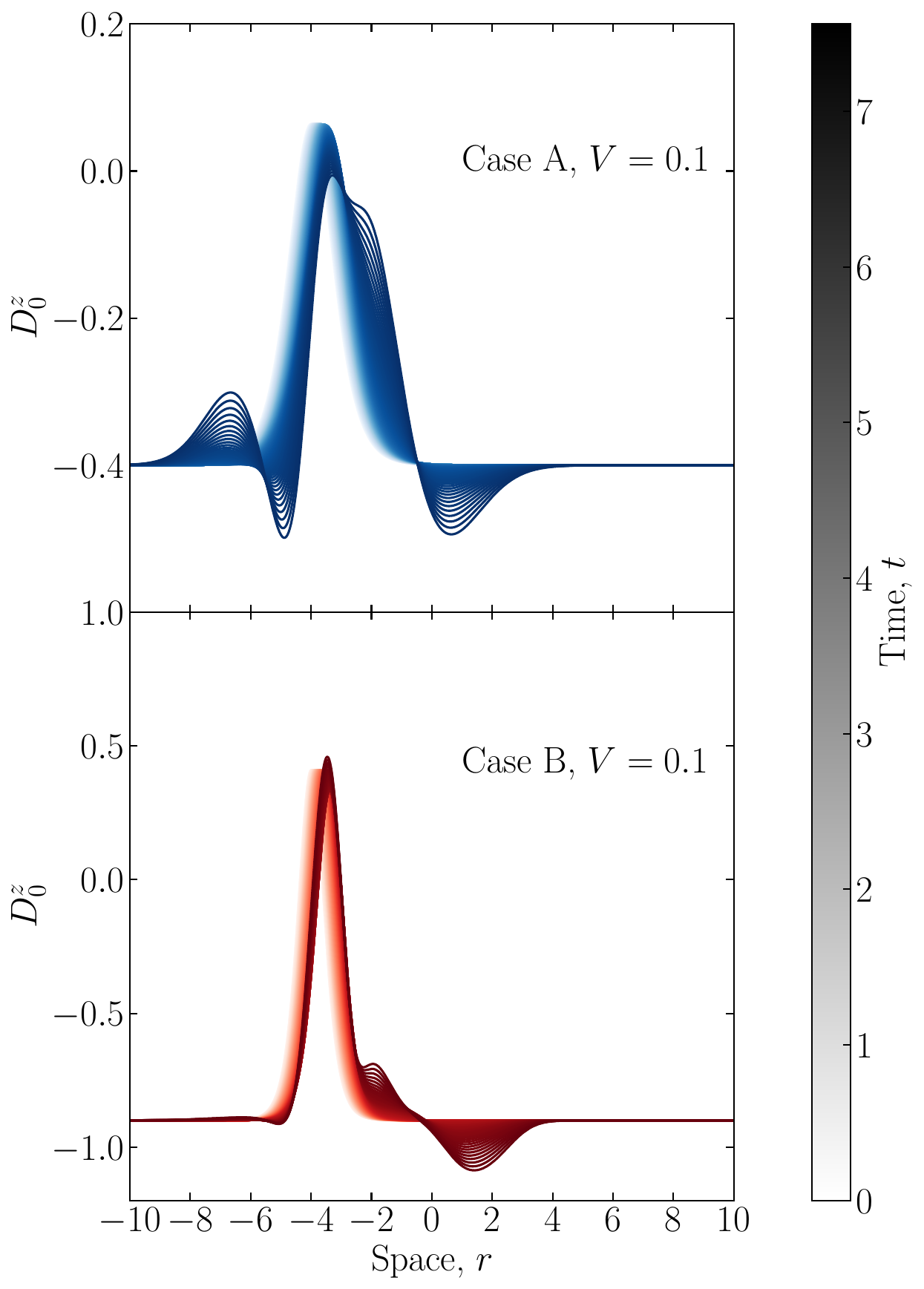}
    \caption{Evolution of an initial soliton with velocity $V=0.1$. The color intensity denotes time.}
    \label{fig:single_soliton_evolution}
    \vskip-6pt
\end{figure}

Figure~\ref{fig:single_soliton_evolution} shows the evolution of the soliton for the three-beam cases A and B introduced in Table~\ref{tab:examples}. In both cases, up to about $t\simeq 3$, the soliton moves indeed with a uniform speed $V=0.1$. We emphasize that this is not a byproduct of our approach, since the only information introduced in the solution of the equations is the initial snapshot of the soliton at $t=0$. The uniform motion comes here because the analytical soliton profile derived in Secs.~\ref{sec:Temporal} and~\ref{sec:Spatial} is an exact solution to the EOMs.

At $t\gtrsim 3$, we see the soliton breaking down at multiple points. Because of the coarseness of the numerical grid, the tail of the soliton acts as a small perturbation to the homogeneous background, which therefore develops perturbation growth; and in turn the center of the soliton, which is also not precisely reproduced by the numerical grid, also rapidly break under perturbation growth. Notice that how coarse the numerical grid is plays little role in how fast the soliton breaks down; the timescale is set mostly by the unstable frequencies of the soliton itself. These are entirely determined by linear stability analysis around the soliton, which however we do not attempt as mentioned earlier. The step of the numerical grid acts here mostly as an effective amplitude of the initial perturbation, and in view of the exponential perturbation growth, it affects the timescale for soliton destruction only logarithmically.

\section{Effect of Matter}

\label{sec:MatterEffects}

In our entire discussion we have ignored the presence of matter that would produce an external refractive effect on neutrinos. There exists a certain perception that the matter effect can be removed by coordinate transformations in flavor or coordinate space, but we will presently see that this perception is deceiving.

We assume that the matter background is homogeneous and that in the laboratory frame, there is a current in the same direction as the symmetry axis of the neutrino gas. In a supernova, the laboratory frame would be the one of a distant observer (Euler coordinates), whereas the frame comoving with the medium are represented by the hydrodynamical Lagrange coordinates. One could use these or any other coordinate frames to study neutrino flavor evolution. In the presence of matter, the EOMs~\eqref{eq:EOM-2} become \cite{Izaguirre:2016gsx}
\begin{equation}
   \bigl(\partial_t+v\partial_r\bigr)\bD_v
   =\bigl[\bigl(\bD_0+\bLam_0\bigr)-v\bigl(\bD_1+\bLam_1\bigr)\bigr]\times\bD_v,
\end{equation}
where $\bLam_0$ is the Bloch vector in the $z$-direction that represents the usual refractive effect of homogeneous matter, and normalized in the same way as the neutrino refractive effect represented by $\bD_0$. If the homogeneous medium moves with velocity $V_\mathrm{mat}$ along the symmetry axis, it contributes a flux term $\bLam_1=V_\mathrm{mat}\bLam_0$ analogous to the neutrino flux term $\bD_1$.

We can certainly study the matter effect in a frame comoving with matter, in which $\bLam_1=0$. If \textit{in this frame} the initial perturbation of the neutrino gas is chosen homogeneous, the EOMs become
\begin{equation}
    \dot{\bD}_v=(\bLam_0+\bD_0)\times \bD_v-v\bD_1\times\bD_v.
\end{equation}
Since $\bLam_0$ is homogeneous in all frames, we may remove it, in analogy to $\bD_0$, by a corotation in flavor space. Therefore, the effect of matter can be entirely eliminated, and the homogeneous temporal solitons are still an exact solution of the EOMs, provided that perturbations are homogeneous in the frame comoving with matter. 

This remark applies in general; the exact pendulum solutions for a homogeneous neutrino gas, which have been thoroughly studied in the literature, are only valid if homogeneity holds in a frame comoving with matter. Remarkably, in the laboratory frame, such solutions are not homogeneous, but actually correspond to our superluminal soliton, moving with a speed $V_\mathrm{mat}^{-1}$. If the motion of matter is nonrelativistic, with $V_\mathrm{mat}\ll1$, the subluminal soliton of course becomes closer and closer to the homogeneous pendulum swing, since the width of the soliton is of order $(\Gamma V_\mathrm{mat})^{-1}$, where $\Gamma$ is the imaginary part of the unstable frequency.

Finally, let us comment on the impact of matter on the static solitons. Due to reciprocity between space and time, here it is $\bLam_1$ which can be easily eliminated by going to a frame (in flavor space) that corotates in space rather than in time. However, differently from the temporal case, the effect of $\bLam_0$ can never be eliminated, since there is no reference frame in which $\bLam_0=0$ if $\bLam_1\neq 0$. After removing the effect of $\bLam_1$, the EOMs take the form
\begin{equation}
    v\partial_r\bD_v=(\bLam_0+\bD_0)\times\bD_v-v\bD_1\times\bD_v.
\end{equation}
This EOM implies that
\begin{equation}
\partial_r \bD_1=\bLam_0\times\bD_0,
\end{equation}
which means that, differently from the case where matter is absent, $\bD_1$ is not conserved. Therefore, the integrability of the model is lost and there is no reason to expect regular solutions.

\begin{figure}[ht]
 \vskip-12pt
    \centering
    \includegraphics[width=1.0\columnwidth]{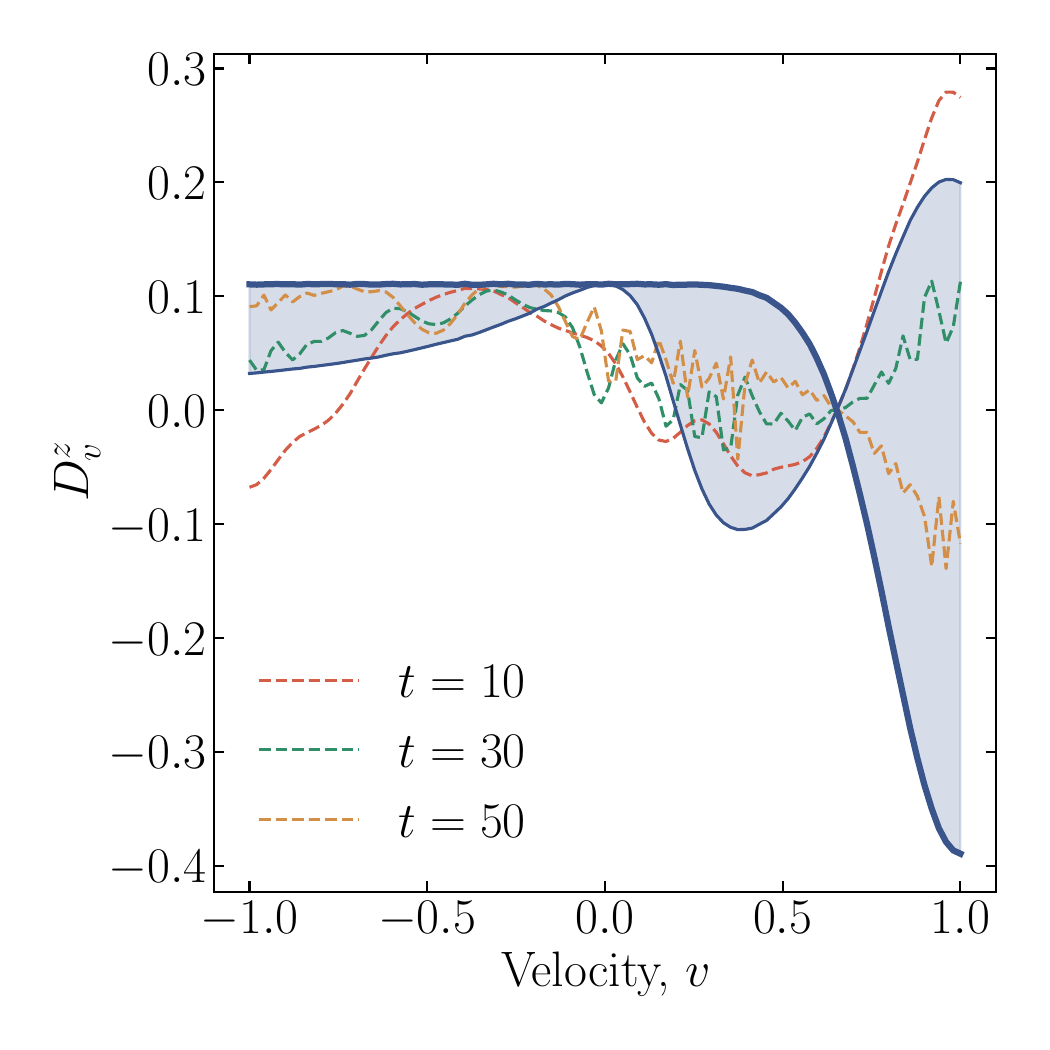}
    \caption{Evolution of $D^z_v$ for the spectrum $G_v$ (thick solid line)      corresponding to Case~D of Ref.~\cite{Padilla-Gay:2021haz}. The undisturbed pendulum motion oscillates between the two solid blue lines, sweeping the shaded region. The time to reach maximum excursion in this case is $t=5.66$, and depends on the chosen initial seed. With matter ($\Lambda_1=2$), the motion is no longer periodic and we show a few snapshots at the indicated~$t$. The numerical resolution has $N=100$ angular modes.}
    \label{fig:example_matter}
    \vskip-6pt
\end{figure}

To illustrate the effect of matter, we perform a numerical experiment on a homogeneous pendulum, assuming a continuous set of beams with the spectrum $G_v$ corresponding to Case D of Ref.~\cite{Padilla-Gay:2021haz}. We show $G_v$ as a thick solid line in Fig.~\ref{fig:example_matter} and, for the undisturbed pendulum, the maximum excursion which can be expressed analytically as discussed in Appendix~\ref{sec:flavor_pendulum} in Eq.~\eqref{eq:maximum-excursion}. For this example, the first moments of the distribution are $G_0=+4.7334$ and $G_1=-5.2665$, whereas the pendulum parameters are $\Omega=1.0743 \pm 1.1121\,i$, corresponding to the natural frequency $\lambda=|\Omega|=1.5462$ and spin parameter $\sigma={\rm Re}\,\Omega/|\Omega|=0.6948$.

Next we include a matter flux $\Lambda_1=2$, which is moderate compared with the moments of the spectrum which measure the refractive effect caused by the neutrinos themselves. To evaluate the integrability of the model, we determine the effective number of degrees of freedom which  describe the motion by computing the non-vanishing eigenvalues of the Gram matrix \cite{Raffelt:2011yb}
\begin{equation}
    {\cal G}_{ij}=\int dt\, \bD_{v_i}(t)\cdot \bD_{v_j}(t),
\end{equation}
where the integral is taken over an arbitrary but numerically suitable period, in our case explicitly the interval $[0,10]$. Our numerical realization uses $N=100$ equidistant beams, and therefore $i,j=1,\ldots,N$. For vanishing $\Lambda_1$ we find three large eigenvalues and all the others much smaller, corresponding to the reduced three-beam nature of the pendulum motion.

For $\Lambda_1=2$, instead, there is no sharp transition between large and small eigenvalues, but roughly there are 20--25 significant large eigenvalues. This result does not depend on our chosen numerical resolution and means that the number of independent modes is much smaller than the number of beams. In other words, the system will not ergodically fill the entire phase space, but rather stay on a lower-dimensional surface than defined by the number of degrees of freedom.

The spectrum of flavor conversion, represented by $D_v^z(t)$, is no longer periodic but shows a much more involuted structure at later times as seen in the snapshots shown in Fig.~\ref{fig:example_matter}. It develops a finer-grained structure, which highlight the larger number of degrees of freedom involved in the motion. On the other hand, even after a long time, the spectrum looks qualitatively similar in that large structures persist as well as large oscillations. The angular structure does not decohere into an ever more fine-grained one. 
In this sense, the motion retains strong collective characteristics
as suggested by the Gram-matrix test. The system no longer moves like a pendulum, but still collectively, in this case corresponding to roughly 20--25 independent degrees of freedom instead of the number $N$ of discrete bins. A systematic study of matter effects and its impact on collective motion is beyond our present ambition.

Similar conclusions hold of course for subluminal solitons, since in their comoving frame the term $\bLam_0$ never vanishes. We conclude that in the presence of matter, static and subluminal solitons no longer exist as exact solutions and would not survive had they been set up as an initial condition.
On the other hand, to which degree collective motion persists or decoherence to ever smaller
angular scales takes place remains to be studied.

On the level of a linear normal-mode analysis, one can always eliminate the effect of matter by redefining the meaning of the frequency and wavenumber of a given perturbation \cite{Izaguirre:2016gsx}. In the nonlinear regime, this is not generally possible so that matter cannot be ignored for the nonlinear evolution of flavor waves.

\section{Conclusion}

\label{sec:Conclusion}

Starting from the fast-flavor pendulum, we have introduced the notion of a soliton, corresponding to the limiting case of a pendulum with a vanishing seed. This solution corresponds to a one-swing pendulum that approaches its unstable fixed point at $t\to\pm\infty$. The existence of such a solution follows from a complex eigenfrequency of the linearized EOMs that fully determines the soliton, for which we have provided an explicit analytical expression. For the formal soliton solution, one does not need to worry about initial conditions. On the other hand, the time when it happens on the infinite time axis is an arbitrary parameter.

A crossed angle spectrum of the neutrino modes is a well-known necessary condition for an instability (or now we would say: for the existence of a soliton). Inspired by the Nyquist criterion, we have derived an additional sufficient condition that can be evaluated without solving the dispersion relation.

Whenever these conditions are satisfied, there also exists a spatial soliton in which the flavor configuration is static and evolves through space similar to the temporal soliton, or like a one-swing ``spatial pendulum'' that approaches its asymptotic state at spatial infinity.

However, our main result is that the temporal and spatial solitons are only limiting cases of more general classes of solutions that can be obtained with the help of Lorentz transformations. One class is that of superluminal solitons, i.e., localized regions in which the neutrino density matrix has off-diagonal coherence, which move with superluminal speed. At infinite speed, the width of this region becomes infinite, and this solution represents the temporal soliton. Likewise, the class of subluminal solitons connects to the spatial soliton in analogous ways.

The seeming violation of causality in the superluminal motion actually derives from the initial conditions, which require correlations to be set up over large scales. In this sense, superluminal solitons highlight in the clearest way the limitations of the homogeneous neutrino gas, which is not truly representative of a realistic setting, since it requires perturbations to be correlated on all scales.

The static soliton, and its subluminal siblings, come closest to the usual picture of a soliton wave, such as the traditional Korteweg-de Vries soliton. However, spatial flavor solitons are extremely fragile in that they break up under small-scale perturbations that are always present and unavoidable in a numerical representation. We have not attempted a full analytical study of their stability, since even simple arguments based on the asymptotic state reveal that they must be as unstable as dictated by the imaginary part of the original linear eigenfrequency that is needed for the soliton to exist in the first place. Therefore, subluminal flavor solitons carry in them their own seed of destruction.

Finally, we clarify the impact of matter effects on all of these solutions. It has become a folk wisdom that in a linearized normal-mode analysis, matter effects can be ``rotated away'' by going to a suitable frame in flavor space. However, the option of eliminating the matter term is much more restricted in the nonlinear regime.

The homogeneous neutrino gas is indeed unaffected by matter, meaning that it still exhibits the temporal soliton (or the fast-flavor pendulum). However, this is only true if the matter flux vanishes. This means that there is only one Lorentz frame in which pendular oscillations can be a valid solution, the frame comoving with matter. Therefore, if matter moves in the laboratory frame with a speed $V_\mathrm{mat}$, the only surviving solution among the class of superluminal solitons is the one moving with a speed $V_\mathrm{mat}^{-1}$. Of course, such a solution requires fine-tuned initial conditions which are homogeneous in the frame comoving with matter. This requirement shows yet another face of the limitations of the homogeneity assumption.

On the other hand, we find that the static and subluminal solitons are always affected by matter, irrespective what frame is chosen, showing that the formal symmetry between space and time is broken by the matter background. In the presence of matter, there is no spatial flavor pendulum or soliton.

The origin of this difference is that even our nominally one-dimensional system involves neutrinos flowing in all zenith-angle directions. Therefore, as a function of spatial coordinate $r$, the phase accrued over some distance $dr$ depends on the actual distance travelled and thus on the zenith angle. Even in the most symmetric configuration, the transverse directions still show up in subtle ways. In this sense, the very existence of soliton solutions once more carries their own cause of destruction, in the case of subluminal solitons in the form of their sensitivity to matter effects. One take-home insight could be to pay more careful attention to matter effects in the context of fast-flavor conversion studies.

Ultimately, it seems that these soliton solutions may not correspond to viable forms of flavor propagation in a real neutrino gas in a real supernova. They are of a more ephemeral, purely mathematical nature, yet they possess a captivating charm that is difficult to resist. In addition, they reveal a lot of unexpected structure in the underlying equations.

\section*{Acknowledgements}

GR acknowledges support by the German Research Foundation (DFG) through the Collaborative Research Centre ``Neutrinos and Dark Matter in Astro and Particle Physics (NDM),'' Grant SFB-1258, and under Germany's Excellence Strategy through the Cluster of Excellence ORIGINS EXC-2094-390783311. DFGF is supported by the Villum Fonden under Project No.\ 29388 and the European Union's Horizon 2020 Research and Innovation Program under the Marie Sk{\l}odowska-Curie Grant Agreement No.\ 847523 ``INTERACTIONS.''

\appendix{}

\section{Instabilities for Single-Crossed Spectra}

\label{sec:SingleCrossed}

\subsection{Nyquist Criterion for Homogeneous Case}

The existence of the temporal soliton is guaranteed by the existence of an unstable eigenmode, namely of a complex solution to the dispersion relation Eq.~\eqref{eq:dispersion_time}. Here we investigate the conditions under which a continuous distribution $G_v$ of velocities $v$ may support an unstable homogeneous
eigenmode. We focus on single-crossed distributions and show that at most one unstable eigenmode may appear, and identify a simple condition for its appearance
based on the Nyquist criterion \cite{nyquist1932regeneration}.

By absorbing $G_0$ in $\Omega$, we may schematically write the dispersion relation as
\begin{equation}
    \tilde{\phi}(\Omega)=\int \frac{v G_v}{vG_1+\Omega}\,dv=0,
\end{equation}
where $G_1=\int dv\,vG_v$ as in the main text. In our previous paper~\cite{Fiorillo:2023mze}, we have discussed how to deal with the pole $v=-\Omega/G_1$ when $\Omega$ is real. Since we are interested in unstable frequencies only, we choose to define the function on the real axis as
\begin{eqnarray}
    \phi(\Omega)&=&\int \frac{v G_v}{vG_1+\Omega+i\epsilon}\,dv
    \nonumber\\
    &=& \int_{\mathrm{PV}}\frac{vG_v}{vG_1+\Omega}\,dv+i\frac{\pi \Omega}{G_1^2}G_{-\Omega/G_1}.
\end{eqnarray}
As we have shown in Ref.~\cite{Fiorillo:2023mze}, this dispersion relation admits the same unstable frequencies. In other words the zeros of $\phi(\Omega)$ and $\tilde{\phi}(\Omega)$ in the upper half of the complex plane defined by Im$(\Omega)>0$ are the same.

The number of unstable frequencies corresponds to the number of zeros of $\phi(\Omega)$ in the upper-half plane. This can be counted by the Nyquist criterion~\cite{nyquist1932regeneration}, which connects the number of zeros to the number of times that the function $\phi(\Omega)$ wraps around the origin in the complex plane as $\Omega$ moves along the real axis. 
Indeed, we have shown in Ref.~\cite{Fiorillo:2023mze} that the dispersion relation for fast modes is analogous in form to the dispersion relation for unstable modes in a collisionless plasma: the function $\phi(\Omega)$ plays here the role of the longitudinal dielectric function, whose zeros correspond to the plasmon modes. The Nyquist criterion must be slightly modified in our context, because the function $\phi(\Omega)$ vanishes as $|\Omega|\to\infty$ only as $\Omega^{-1}$, which precludes a naive application of the zero-counting theorem. Therefore, we here derive the result from scratch.

To count the number of unstable frequencies, we consider the integral
\begin{equation}
    I=\int_{-\infty}^{+\infty} \frac{d\log\phi(\Omega)}{d\Omega}\,d\Omega
\end{equation}
taken over the real axis. The integral can be closed by a semicircle in the upper half complex plane. The integral on the semicircle does not vanish, because the function $\phi(\Omega)\to-G_1/\Omega$, and therefore the integral on the semicircle is
\begin{equation}
    J=-\int \frac{d\Omega}{\Omega}=-i\pi.
\end{equation}
The integral over the entire closed contour is equal to the number of poles of the integrand function in the upper half plane. 

Assuming that $\phi(\Omega)$ does not have poles itself in this region, these poles are simply the zeros of $\phi(\Omega)$. From each of these zeros, the integral draws a contribution $2\pi i$. Therefore, it follows that
\begin{equation}
    I+J=2\pi i N,
\end{equation}
where $N$ is the number of complex unstable frequencies. From here it follows
\begin{equation}\label{eq:phase_acquired}
    \phi(\Omega\to+\infty)=\phi(\Omega\to -\infty) e^{i\pi} e^{2i\pi N}.
\end{equation}
Therefore, as $\Omega$ grows from $-\infty$ to $+\infty$, the function $\phi(\Omega)$ starts from $0$ and returns to $0$ after having changed its sign and wrapped itself around the origin $N$ times.

\begin{figure}[t!]
    \centering
    \includegraphics[width=0.490\columnwidth]{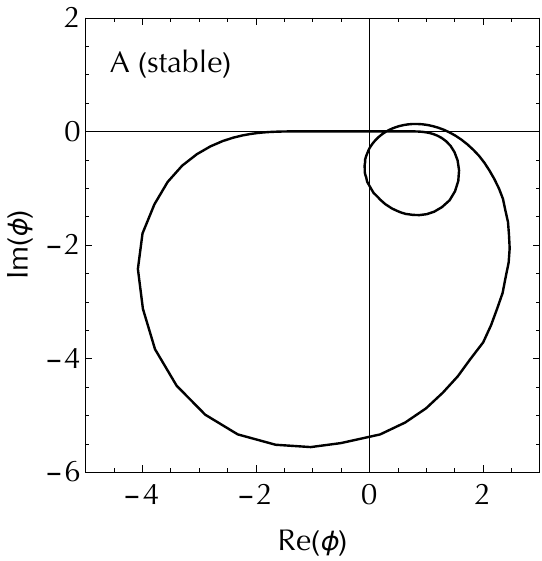}
    \includegraphics[width=0.496\columnwidth]{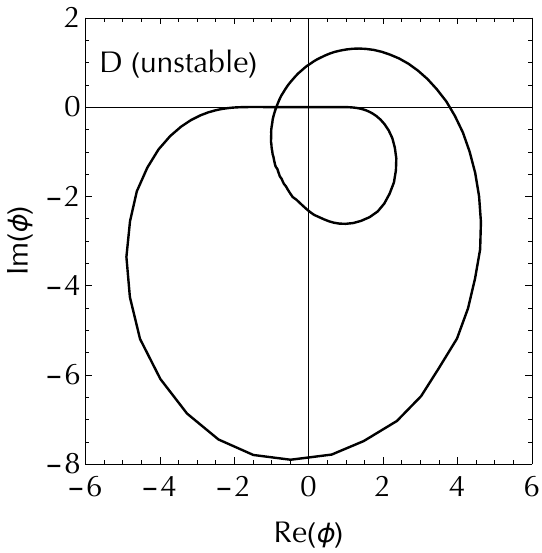}
    \caption{Nyquist diagrams for the angular distribution of case A (left)
     and case D (right) of Ref.~\cite{Padilla-Gay:2021haz}. We show the trajectory of the function $\phi(\Omega)$ in the complex plane, as $\Omega$ evolves from $-\infty$ to $+\infty$ along the real axis. Since the angular distribution has non-analytical discontinuities at $v=-1$ and $v=+1$, we smooth them out introducing a rapid cutoff at these positions.}
    \label{fig:nyquist_diagrams}
\end{figure}

We first confirm the validity of the criterion empirically in two simple cases, the angular distributions for cases~A and~D of Ref.~\cite{Padilla-Gay:2021haz}. For both cases, we adjust the sign of the distribution such that $G_1>0$. Figure~\ref{fig:nyquist_diagrams} shows the trajectories drawn by $\phi(\Omega)$ in the complex plane as $\Omega$ runs from $-\infty$ to $+\infty$. At $\Omega\to-\infty$, $\phi(\Omega)\to G_1/\Omega$, so that $\phi$ starts from $0$ and moves along the negative real axis. At $\Omega\to+\infty$, $\phi(\Omega)$ returns again to the origin from the positive real axis, therefore confirming our result in Eq.~\eqref{eq:phase_acquired} that it acquires a phase of~$\pi$. 

In addition, for case D (right panel), the function $\phi(\Omega)$ wraps completely around the origin once, corresponding to one unstable mode. Indeed, Ref.~\cite{Padilla-Gay:2021haz} finds this distribution to be unstable. On the other hand, for case A (left panel) the trajectory of $\phi(\Omega)$ does not wrap around the origin, corresponding to no unstable mode, as also found in Ref.~\cite{Padilla-Gay:2021haz}.

We can now derive a simple criterion to determine whether a single-crossed spectrum $G_v$ admits an instability without explicitly solving the dispersion relation. The key idea is that the function $\phi(\Omega)$ can only cross the real axis, for real $\Omega$, either at $\Omega=0$ or at $\Omega=-\vc G_1$, where $\vc$ is the position of the crossing $G_{\vc}=0$. Without loss of generality, we choose $G_1>0$. Therefore, at $\Omega\to -\infty$, the function $\phi(\Omega)\to G_1/\Omega$ starts from $0$ tangent to the negative real axis. At $\Omega\to+\infty$, $\phi(\Omega)$ returns to $0$ tangent to the positive real axis. The number of times that it wraps around the origin in between corresponds to the number of unstable frequencies. In order to wrap around the origin once, $\phi(\Omega)$ must cross the real axis first on the positive side, as in both panels of Fig.~\ref{fig:nyquist_diagrams}. Afterwards, to surround the origin, it should cross it again on the negative side, as in the right panel of Fig.~\ref{fig:nyquist_diagrams}, so that it can finally return to the positive side and asymptotically reach zero. It is easy to see that with one crossing, $\phi(\Omega)$ can wrap around the origin at most once, so there is at most one unstable frequency.

Therefore, the criterion for the existence of an instability is the following: starting from the negative real axis, the function $\phi(\Omega)$ should cross first the positive real semi-axis, and then the negative one.  We can now state our criterion more plainly: If $\vc>0$, we have an instability if $\phi(-\vc G_1)>0$ and $\phi(0)=G_0/G_1<0$. If $\vc<0$, we have an instability if $\phi(0)>0$ and $\phi(-\vc G_1)<0$. This criterion allows us to identify the presence of an instability for single-crossed spectra without explicitly solving the dispersion relations themselves. Notice that, while in the derivation we assumed $G_1>0$, the final result is independent of this assumption.

\subsection{Superluminal Soliton}

The criterion for the existence of a homogeneous instability can be immediately extended to a criterion for the existence of superluminal solitons with speed $v_\mathrm{soliton}=V^{-1}$. In Sec.~\ref{sec:superluminal}, we have shown that the EOMs for the superluminal solitons are identical to the EOMs of the homogeneous system in a boosted frame. The dispersion relation can therefore still be written as
\begin{equation}
    \phi(\Omega')=\int \frac{v' S^{z,0}_{v'}}{v' G^{\prime}_1+\Omega'}d\xi'=0.
\end{equation}
Notice that $S^{z,0}_{v'}dv'=\left[\omega' G_v dv\right]_{v'=k'/\omega'}$ by definition, evaluated at the asymptotic condition in terms of $G_v$, and $\omega'=(1-vV)/(1-V^2)^{1/2}$. Here, as in the main text, $G^\prime_0=\gamma(G_0-VG_1)=\sum_{v'} S^{z,0}_{v'}$, $G^\prime_1=\gamma(G_1-VG_0)=\sum_{v'} S^{z,0}_{v'} v'$, evaluated at the asymptotic condition.

To extend our previous criterion, the crossing point in the new frame is
$v'_{\rm c}=(\vc-V)/(1-\vc V)$. If $v'_{\rm c}>0$, an instability is present if $\phi(0)<0$ and $\phi(-v' D^{\prime}_1)>0$, and vice versa if $v'_c<0$. Performing the appropriate replacements, we find that the instability is present if
\begin{equation}
    \vc>V,\;\frac{G_0-V G_1}{G_1-V G_0}<0,\; \int \frac{(v-V)(1-v V)G_v}{(G_1-V G_0)(v-\vc)}\,dv>0
\end{equation}
or
\begin{equation}
    \vc<V,\;\frac{G_0-V G_1}{G_1-V G_0}>0,\; \int \frac{(v-V)(1-v V)G_v}{(G_1-V G_0)(v-\vc)}\,dv<0.
\end{equation}
The two criteria may be more compactly combined as 
\begin{subequations}
    \begin{eqnarray}
    \frac{(\vc-V)(G_0-VG_1)}{G_1-VG_0}&<&0,
    \\[1ex]
       \int \frac{(v-V)(1-v V)G_v}{(G_0-V G_1)(v-\vc)}\,dv&<&0,
    \end{eqnarray}
\end{subequations}
which both need to be satisfied for an instability to exist.

Notice that, for $V=\vc$, the trajectory of $\phi(\Omega')$ in the Nyquist diagram can touch the real axis only at one point (except the trivial tangent behavior at $\Omega\to \pm\infty$), since the two crossings at $v=\vc$ and $v=V$ merge into a single point of tangency to the real axis, which does not allow the trajectory to wrap around the origin. Therefore, superluminal solitons with $V=\vc$ are not supported.

\subsection{Static and Subluminal Soliton}

For the subluminal and static soliton, the existence of an unstable frequency needs not be studied separately. In fact, the dispersion relation for a superluminal soliton can be written in terms of the primed quantity as
\begin{equation}\label{eq:a11}
\int \frac{k'}{v' G'_1-G'_0+\Omega'} \left[G_v dv\right]_{k'=\gamma(v-V),\,v'=\frac{v-V}{1-v V}}=0.
\end{equation}
The corresponding dispersion relation is
\begin{equation}\label{eq:a12}
\int\frac{k'}{v' G'_1-G'_0+v' K'}\! \left[G_v dv\right]_{k'=\gamma(v-V),\,v'=\frac{v-V}{1-v V}}=0.
\end{equation}
From here, we deduce that if a superluminal soliton is supported with speed $v_\mathrm{soliton}=V^{-1}$ and frequency $\Omega'$, there is also a corresponding solution for a subluminal soliton with speed $V$, whose wavenumber $K'$ is obtained from
\begin{equation}\label{eq:a13}
    \frac{G'_0-\Omega'}{G'_1}=\frac{G'_0}{G'_1+K'}.
\end{equation}
This establishes a dual relation among superluminal and subluminal solitons. The relation may be directly expressed in terms of the moments of the original spectrum $G_v$, giving
\begin{equation}\label{eq:dual_solitons}
    \frac{G_0-V G_1 -\Omega}{G_1-V G_0}=\frac{G_0-V G_1}{G_1-VG_0+K},
\end{equation}
where $K=K'/\gamma$ and $\Omega=\Omega'/\gamma$. In particular, this means that if a single-crossed spectrum $G_v$ permits a homogeneous temporal soliton, characterized by the linear eigenfrequency $\Omega$, then there also exists a static spatial soliton corresponding to the linear wavenumber
\begin{equation}
  K=\frac{G_1 \Omega}{G_0-\Omega}  
\end{equation}
in the limit $V=0$ of Eq.~\eqref{eq:dual_solitons}. 

Notice that, while we have used the integral notation in Eqs.~\eqref{eq:a11} and~\eqref{eq:a12}, the final results Eqs.~\eqref{eq:a13} and~\eqref{eq:dual_solitons} are valid for a discrete number of beams as well, since they only involve the redefinition in the denominators of Eqs.~\eqref{eq:a11} and~\eqref{eq:a12}. In other words, any single-crossed spectrum, discrete or continuous, that permits a homogeneous temporal soliton also permits a static spatial one.

%%%%%%%%%%%%%%%%%%%%%%%%%%%%%%%%%%%%%%%%%%%%%%%%%%%%%%%%%%%%%%%%%%%%%%
\section{Gyroscopic Pendulum}                 \label{sec:gyropendulum}
%%%%%%%%%%%%%%%%%%%%%%%%%%%%%%%%%%%%%%%%%%%%%%%%%%%%%%%%%%%%%%%%%%%%%%

\subsection{Equations of Motion}

We briefly review the gyroscopic pendulum, following Appendix~B of Ref.~\cite{Raffelt:2011yb}. This contraption, also known as symmetric heavy top, Lagrangian top, or spherical pendulum with spin, is an axially symmetric body, spinning around its symmetry axis (moment of inertia~$I_s$) with support on this axis \cite{Klein-Sommerfeld:2008}. Its moment of inertia relative to that point is $I$, mass $M$, gravitational acceleration $g$ along the \hbox{$z$--direction}, distance $\ell$ between support and center of mass, and angle $\vartheta$ relative to the $z$--direction, i.e., this angle is counted relative to the upward direction. The potential energy is $V=M g\,\ell\,\cos\vartheta$. If the top were essentially point like, one would have $I=M\ell^2$, but in general one considers an extended body, where the center of mass and the ``center of oscillation'' are different.

The angular momentum ${\bf S}$ along the symmetry axis (spin) has kinetic energy $T_{\rm spin}={\bf S}^2/2I_s$. The point of support is on the symmetry axis, preventing a torque to change $S=|{\bf S}|$, and so both $S$ and $T_{\rm spin}$ are conserved. We are only interested in the gyroscope's orbital motion, not the internal spin motion, so that $T_{\rm spin}$ is an additive constant to the Hamiltonian. For given $\bS$, the orbital motion will be the same, irrespective of $T_{\rm spin}$, so that we may assume $I_s=I$ for simplicity.
The entire kinetic energy can then be expressed in terms of the single $I$ as $T=\bJ^2/2I$, where $\bJ$ is the total angular momentum. 

We may further use the radius vector $\bR$ from the point of support to the center of mass as a spatial coordinate. The potential energy provided by gravity is $g M \bG\cdot\bR$ with $|\bR|=\ell$ and the gravitational unit vector $\bG$ is defined to point upward. The Hamiltonian of this overall mechanical system is
\begin{equation}
    H=g M\,\bG\cdot\bR+\frac{\bJ^2}{2I}.
\end{equation}
The equations of motion derive from this Hamiltonian through the Poisson brackets 
$\dot\bR=\{\bR,H\}$ and $\dot\bJ=\{\bJ,H\}$. Notice that $\{J_i,J_j\}=\epsilon_{ijk}J_k$, 
$\{J_i,R_j\}=\epsilon_{ijk}R_k$, and $\{R_i,R_j\}=0$. Explicitly one finds
\begin{equation}\label{eq:gyroscope}
    \dot\bR=\frac{\bJ\times\bR}{I}
    \quad\hbox{and}\quad
    \dot\bJ=g M\bG\times\bR.
\end{equation}
The physical content is that $\bJ$ spawns a differential rotation as behooves the total angular momentum, whereas the vertical force of gravity exerts a torque, proportional to $\bG\times\bR$, that changes the total angular momentum. This form allows one to port the gyroscope EOMs directly to those of the Bloch vectors of the FFC system.

However, to solve the EOMs explicitly, we continue with the mechanical system in the spirit of the textbook literature. The orbital angular momentum is ${\bf L}=I\,{\bf r}\times{\dot{\bf r}}$, where ${\bf r}$ is a unit vector along the symmetry axis. It marks the top's orientation with zenith angle $\vartheta$ and azimuth angle~$\varphi$. The orbital kinetic energy is
\begin{equation}\label{eq:pendulumkinetic}
T_{\rm orb}=\frac{{\bf L}^2}{2I}=
\frac{1}{2}\,I\,\dot{\bf r}^2 =\frac{1}{2}\,I\,
\bigl(\dot\vartheta^2+\dot\varphi^2\sin^2\vartheta\bigr).
\end{equation}
The total angular momentum ${\bf J}={\bf L}+{\bf S}$ has conserved \hbox{$z$--component}, where $S_z=S\,\cos\vartheta$ and $L_z=I\dot\varphi\sin^2\vartheta$. Here one factor of $\sin\vartheta$ comes from the projection of ${\bf r}$ on the transverse plane and the velocity is $\dot\varphi\sin\vartheta$. Therefore, $J_z=I\dot\varphi\sin^2\vartheta+S\cos\vartheta$ is conserved and
\begin{equation}\label{eq:phimotion}
\dot\varphi=\frac{J_z-S\cos\vartheta}{I\sin^2\vartheta}.
\end{equation}
Therefore
\begin{equation}
T_{\rm orb}=\frac{1}{2}\,I\,\dot\vartheta^2+
\frac{(J_z-S\cos\vartheta)^2}{2 I\sin^2\vartheta}
\end{equation}
and the total energy $E=T+V$ is
\begin{equation}\label{eq:pendulumtotal}
E=\frac{I}{2}\,\dot\vartheta^2+\frac{(J_z-S\cos\vartheta)^2}{2 I\sin^2\vartheta}
+M g \ell \cos\vartheta\,.
\end{equation}
This has the form $E=I\,\dot\vartheta^2/2+V(\vartheta)$, where $V(\vartheta)$ is a potential given in terms of conserved quantities fixed by initial conditions.

Next we introduce $c=\cos\vartheta$ as independent variable so that $\dot\vartheta^2=\dot c^2/\sin^2\vartheta$ and find the third-order polynomial that is characteristic for the gyroscopic pendulum
\begin{equation}\label{eq:thetadot3}
\dot c^2=2\,\frac{E-Mg\ell\,c}{I}\,(1-c^2)-\left(\frac{J_z-S\,c}{I}\right)^2.
\end{equation}
When $J_z=S=0$ we have a plane pendulum, where
\begin{equation}
\lambda^2=M g\,\ell/I
\end{equation}
provides the natural frequency $\lambda$. Moreover, we express 
$J_z=j_z S$ and use the dimensionless spin parameter~\cite{Padilla-Gay:2021haz}
\begin{equation}
    \sigma=\frac{S}{2\lambda I}.
\end{equation}
With these parameters, the EOMs are
\begin{subequations}
\begin{eqnarray}
    \kern-2em \dot\varphi&=&2\lambda \sigma\,\frac{j_z-c}{1-c^2},
    \\[1ex]
    \kern-2em \dot c^2&=&4\lambda^2\biggl[\frac{1}{2}\,\bigl(\varepsilon-c\bigr)\,\bigl(1-c^2\bigr)-\sigma^2\bigl(j_z-c\bigr)^2\biggr],
\end{eqnarray}
\end{subequations}
where $\varepsilon=E/I\lambda^2$ is the dimensionless total energy.

\subsection{Soliton Solution}

In the context of the flavor pendulum, we are interested in motions that consist of a precession and a nutation between two limiting zenith angles $0\leq\vartheta_1\leq\vartheta\leq\vartheta_2\leq\pi$, corresponding to $1\geq c_1\geq c\geq c_2\geq-1$. In particular, if the highest position is $c_1=1$, the upright orientation, then $J_z=S$ and $j_z=1$ as well as $\varepsilon=1$. Moreover, we may absorb $\lambda$ in the definition of time, implying
\begin{subequations}
\begin{eqnarray}\label{eq:phi-1}
    \dot\varphi&=&\frac{2\sigma}{1+c},
    \\[1ex]
    \label{eq:z-1}
    \dot c^2&=&4\,(1-c)^2\biggl[\,\frac{1+c}{2}-\sigma^2\,\biggr].
\end{eqnarray}
\end{subequations}
If $\sigma>1$, the second equation is true only for $c=1$ and the pendulum is stuck in the upright ``sleeping top'' position.

Otherwise its lowest point $c_2$ requires $\dot c^2=0$ and thus the bracket in Eq.~\eqref{eq:z-1} must vanish. One thus finds
\begin{equation}
    c_2=\cos\vartheta_{\rm min}=-1+2\sigma^2,
\end{equation}
which indeed varies between $-1\leq c\leq +1$ for $0\leq \sigma\leq 1$. If this lowest point occurs at $t=0$, the solution is
\begin{subequations}\label{eq:soliton-solution}
\begin{eqnarray}\label{eq:phi-2}
    \varphi(t)&=&\sigma\lambda t
    +\arctan\left[\frac{\sqrt{1-\sigma^2}}{\sigma}\,\tau\right],
    \\[1ex]
    \label{eq:z-2}
    c(t)&=&-1+2\left[\sigma^2+(1-\sigma^2)\tau^2\right]
    \\[1ex]
    &=&1-\frac{8\left(1-\sigma^2\right)}{\bigl(e^{\sqrt{1-\sigma^2}\lambda t}
    +e^{-\sqrt{1-\sigma^2}\lambda t}\bigr)^2},
    \nonumber
\end{eqnarray}
\end{subequations}
where we have restored the natural pendulum frequency $\lambda$. Moreover, we have introduced the time coordinate
\begin{equation}
    \tau=\tanh\left[\sqrt{1-\sigma^2}\,\lambda t\right],
\end{equation}
which varies as $-1<\tau<+1$ for $-\infty<t<\infty$.

The solution is a pendulum that swings only once (a~temporal soliton) and at $t\to\pm\infty$ approaches the upright position, the latter being an unstable fixed point. The azimuthal variation includes an overall precession with frequency $\omega_{\rm P}=\sigma\lambda$ that can be removed by going to a corotating frame.

The complete orbital motion of the gyroscope is encoded in the polar angles given in Eq.~\eqref{eq:soliton-solution}. We may write the solution for the unit vector providing the orientation of the gyroscope in the form
\begin{equation}\label{eq:Explicit-r}
    \br(t)=\begin{pmatrix}s(t)\cos\varphi(t)\\
    s(t)\sin\varphi(t)\\
    c(t)
    \end{pmatrix},
\end{equation}
where $s(t)=\sqrt{1-c(t)^2}$. 

In addition, we will later need an explicit solution for $\bJ(t)$ that also follows from these results. The total angular momentum is a sum of the spin and orbital part, $\bJ=\bS+\bL$. The former is $\bS=S\,\br$, where $S=2\sigma\lambda$ is the conserved spin. As discussed earlier, in units where the moment of inertia is taken to be unity, $\bL=\br\times\dot\br$, so that
\begin{equation}\label{eq:Explicit-J}
    \bJ=S\,\br+\br\times\dot\br.
\end{equation}
The derivatives can now be obtained by explicit differentiation of the solutions in Eq.~\eqref{eq:soliton-solution}.

\subsection{Connection to Linear Normal Modes}

Near the upright position, at very early or very late times, the solution can be linearized. However, this makes only sense for the zenith angle $\vartheta$ or $s=\sin\vartheta$, not for $c=\cos\vartheta$, which to lowest order is $c=1$ at early and late times. So we rather look at the $x$-$y$ components of the pendulum vector ${\bf r}$. For very early times ($t\to-\infty$) the linearized solution is explicitly
\begin{subequations}
\begin{eqnarray}\label{eq:phi-3}
    \varphi(t)&=&\sigma\lambda t
    -\varphi_\sigma,
    \\[1ex]
    \label{eq:z-3}
    s(t)&=&4\sqrt{1-\sigma^2}\,\,e^{\sqrt{1-\sigma^2}\lambda t},
\end{eqnarray}
\end{subequations}
where
\begin{equation}
    \varphi_\sigma=\arctan\left[\frac{\sqrt{1-\sigma^2}}{\sigma}\right].
\end{equation}
If we express the pendulum vector ${\bf r}$ in a spherical basis through $r_0=z$ and $r_\pm=x\pm i y$, we may express the linearized solution as $r_0=1$ and
\begin{equation}
    r_\pm=4\sqrt{1-\sigma^2}\,e^{\pm i\varphi_\sigma}\,
    e^{\left(\sqrt{1-\sigma^2}\mp i\sigma\right)\lambda t}.
\end{equation}
The physical content is the same: The pendulum precesses around the $z$-direction with a frequency $\omega_{\rm P}$ and its $x$-$y$ component grows exponentially with a rate $\Gamma$ where
\begin{equation}
    \omega_{\rm P}=\sigma\lambda
    \quad\hbox{and}\quad
    \Gamma=\sqrt{1-\sigma^2}\,\lambda.
\end{equation}
Conversely, the pendulum parameters are \cite{Padilla-Gay:2021haz}
\begin{equation}\label{eq:pendulumpars-App}
    \lambda = \sqrt{\omega_{\rm P}^2+\Gamma^2}
    \quad\hbox{and}\quad
    \sigma = \frac{\omega_{\rm P}}{\sqrt{\omega_{\rm P}^2+\Gamma^2}}
\end{equation}
in terms of the linearized eigenfrequency $\Omega=\omega_{\rm P}\pm i\Gamma$. This complex frequency fully determines the soliton and sometimes we call it the ``soliton frequency.''

\section{Flavor pendulum}\label{sec:flavor_pendulum}

A single-crossed spectrum admits at most one unstable mode, as we discussed in Appendix~\ref{sec:SingleCrossed}. With a single unstable mode, the most general motion that the system can perform is a pendular one. The simplest system that can perform a pendular motion is on consisting of three Bloch vectors. Therefore, we first recall how a three-beam system dynamics is equivalent to a pendulum \cite{Padilla-Gay:2021haz}. We later connect the three-beam system with a generic single-crossed, unstable spectrum.

\subsection{From Three Beams to Pendulum and Soliton}

The FFC system does not involve any external vector, unlike the slow system, where the mass direction is singled out. Therefore, to mimic a gyroscopic pendulum, a FFC system $\{v,\bD_v\}$ requires at least three beams so that different linear combinations can play the role of $\bG$, $\bR$, and $\bJ$. Any FFC system fulfilling the homogeneous EOMs of Eq.~\eqref{eq:homogeneous} has the conserved vector $\bD_0$ and the vector $\bD_1$ with conserved length. Moreover, for a three-mode system one can define \cite{Padilla-Gay:2021haz}
\begin{equation}\label{eq:J-expression}
    \bJ=\bD_2-v_s\bD_1=\sum_{i=1}^3\left(v_i-v_s\right)v_i\bD_{v_i},
\end{equation}
where $v_s=v_1+v_2+v_3$. These Bloch vectors obey
\begin{subequations}
    \begin{eqnarray}
        \dot\bD_0&=&0,\\
        \dot\bD_1&=&\bJ\times\bD_1,
        \label{eq:rotation}\\
        \dot\bJ&=&v_1v_2v_3\bD_0\times\bD_1.
    \end{eqnarray}
\end{subequations}
For any three-mode system with $\bD_0\not=0$, $\bD_1\not=0$, and $v_1v_2v_3\not=0$, these EOMs are equivalent to Eq.~\eqref{eq:gyroscope} of a gyroscope. Here we have already absorbed the neutrino-neutrino interaction energy $\mu$ in the definition of dimensionless time, but otherwise we recognize from Eq.~\eqref{eq:rotation} that $\mu$ is equivalent to $I^{-1}$, the inverse moment of inertia of the gyroscope.

The conserved vector $\bD_0$ defines the $z$-axis, although not necessarily its sign. To obtain a soliton we turn to a more restricted system, where all three $\bD_v(t)$ become asymptotically collinear at $t\to\pm\infty$, i.e., the system asymptotically approaches the ``sleeping top'' unstable fixed point. This system is characterized by $\{v,G_v\}$ with three discrete velocities and the spectrum $G_v$ conditional on $v_1v_2v_3 G_0 G_1\not=0$. Because $J_z$ is conserved and is identical with the spin $S$ in the upright pendulum position, one finds
\begin{equation}
    S=G_2-(v_1+v_2+v_3)\,G_1.
\end{equation}
The dispersion relation Eq.~\eqref{eq:dispersion_time}, after absorbing $G_0$ in $\Omega$, then implies the eigenfrequency
\begin{equation}
 \Omega=\frac{S}{2}\pm i\sqrt{v_1v_2v_3G_0G_1-\left(\frac{S}{2}\right)^2}
\end{equation}
as stated in Eq.~\eqref{eq:three-mode-frequency} of the main text. This eigenfrequency
represents two complex conjugate solutions (and thus a true ``soliton frequency'') only if the argument of the square root is positive, which also implies that the first term under the square root must be positive. In this case follows the identification of the pendulum natural frequency
$\lambda=\sqrt{v_1v_2v_3 G_0G_1}$ and spin parameter $\sigma=S/2\lambda$ stated in Eq.~\eqref{eq:pendulumpars} of the main text.

\subsection{From Soliton to Three Beams}

We may reverse this problem and ask for a three-beam realization of a soliton with given frequency $\Omega=\wP\pm i\Gamma$. For example, a given supernova-inspired single-crossed continuous spectrum $G_v$ such as those used in Ref.~\cite{Padilla-Gay:2021haz} may provide us with the corresponding $\Omega$ and we may wish to construct an equivalent three-beam system. As a first step, one derives the equivalent natural pendulum frequency and spin parameter according to Eq.~\eqref{eq:pendulumpars-App}. 

One immediately identifies $S=J_z=2\wP=2\,{\rm Re}\,\Omega$. Moreover, from Eq.~\eqref{eq:three-mode-pendulum-pars} we recall that $\lambda^2=\wP^2+\Gamma^2=v_1v_2v_3 G_0 G_1$ so that overall the connections are
\begin{subequations}
    \begin{eqnarray}
        2\,{\rm Re}\,\Omega&=&G_2-(v_1+v_2+v_3)\,G_1\\
        \label{eq:three-pars-2}
        |\Omega|^2&=&v_1v_2v_3 G_0 G_1.
    \end{eqnarray}
\end{subequations}
Therefore, the remaining five three-mode parameters are very degenerate. We may pick $-1\leq v_1<v_2<v_3\leq+1$ and $G_0$ and $G_1$ anyway we like such that their product provides the desired $|\Omega|^2$.

After these choices have been made, the spectrum $G_{v_i}$ is provided explicitly by Eq.~(S16) of Ref.~\cite{Padilla-Gay:2021haz}, which can be written as
\begin{equation}
    G_{v_i}=\frac{v_i S+v_i^2 G_1+v_1v_2v_3 G_0}{v_i \prod_{j\not=i}(v_i-v_j)}
\end{equation}
for each $i=1$, 2 and 3.

One may wonder if this construction indeed yields a single-crossed spectrum as it must, i.e., three $G_{v_i}$ that do not all have the same sign. We observe that this equation may be rewritten as
\begin{equation}
    G_{v_i}=\frac{\Gamma^2+\left(G_0 \prod_{j\neq i} v_j + \wP\right)^2}{G_0 \prod_{j\neq i} v_j \prod_{j\neq i} (v_j-v_i)}.
\end{equation}
Without loss of generality, we can choose $G_0>0$. The signs then depend on the two products in the denominator. The $\prod_{j\neq i} (v_j-v_i)$ is positive for $G_{v_1}$ and $G_{v_3}$ and negative for $G_{v_2}$. If we want no crossing, we would need $v_1 v_3<0$, $v_1 v_2>0$, and $v_2 v_3>0$. This is not possible for any three numbers, so at least one crossing will appear.

If the original starting point was a continuous spectrum, we may wish to use the same moments $G_0$ and $G_1$ for the three-mode realization. Notice, however, that it is not necessarily assured that this is possible. Given $|\Omega|^2$ and after choosing $G_1$ and noticing that $|v_1v_2v_3|<1$, it is not necessarily assured that we can find three velocities such that Eq.~\eqref{eq:three-pars-2} can be satisfied. Or turning this around, for a given continuous unstable $G_v$, it is not obvious that always $|\Omega|^2<|G_0 G_1|$.

For the moment we do not have a mathematical proof that this will always be the case, but extensive empirical searches have not turned up a counter example. It appears that this condition can be saturated in a limiting sense by three-mode examples with all beams having $v=\pm1$ in a limiting sense. So we conjecture that this condition applies for any continuous or discrete $\{v,G_v\}$, implying that the three-beam realization never requires a superluminal beam.

\subsection{Explicit Solution for the Entire Spectrum}

Given a discrete or continuous system $\{v,\bD_v\}$ that supports a temporal soliton implies that $\bD_1$ moves like a pendulum, but also means that all individual $\bD_v$ move in collective ways and each of them returns to its asymptotic position at $t\to+\infty$. Each individual $\bD_v$ follows $\dot\bD_v=(\bD_0-v\bD_1)\times\bD_v$ with $\bD_0$ a conserved vector and $\bD_1$ following the pendulum motion. The behavior of $\bD_v$ is not caused by it being a member of a larger collective ensemble---it only feels $\bD_0$ and $\bD_1$. It would follow a periodic motion even if $\bD_0$ and $\bD_1$ were externally prescribed Bloch vectors as long as $\bD_1$ is prescribed to follow a gyroscope motion. With our tools it is now straightforward to write the motion of a given $\bD_v$ explicitly in terms of the pendulum motion without having to solve the differential equation for~$\bD_v$.

To this end we assume a three-mode system $\{v_i,\bD_{v_i}\}$ with $i=1$, 2 and 3 that supports a soliton.
This three-mode system could be a set of carrier modes (auxiliary spins) of a larger system. We may then express $\bD_0$, $\bD_1$ and $\bJ$ in terms of the three $\bD_{v_i}$ as in Eq.~\eqref{eq:J-expression}, and then conversely the three $\bD_{v_i}$ in terms of $\bD_0$, $\bD_1$ and $\bJ$. We further know that any Lax vector
\begin{equation}\label{eq:Lu}
    \bL_u=\sum_{i=1}^3\frac{v_i\bD_{v_i}}{u-v_i}
\end{equation}
fulfills the original precession equation in the form $\dot\bL_u=(\bD_0-u\bD_1)\times\bL_u$. Inserting the explicit expressions for the three $\bD_{v_i}$ in Eq.~\eqref{eq:Lu} reveals
\begin{equation}
    \bL_u=\frac{v_1v_2v_3\bD_0+u^2\bD_1+u\bJ}{(u-v_1)(u-v_2)(u-v_3)}.
\end{equation}
The denominator is just an arbitrary factor---we are only interested in the orientation of $\bL_u$, whereas its conserved length can be arbitrarily chosen. Moreover, we observe that the natural pendulum frequency is given by $\lambda^2=v_1v_2v_3 G_0 G_1$ where we use the spectrum $G_v=D^z_v(\pm\infty)$.
Therefore, $v_1v_2v_3\bD_0=\lambda^2\,\hat{\bf z}/G_1$. The pendulum motion is $\bD_1(t)=G_1\,\br(t)$, where $\br(t)$ is the explicit solution Eq.~\eqref{eq:Explicit-r}, whereas $\bJ(t)$ was explicitly provided in Eq.~\eqref{eq:Explicit-J}. Therefore, with modified normalization, 
Eq.~\eqref{eq:Lu} can be expressed as
\begin{equation}\label{eq:Lax-Pendulum}
    \bL_u(t)=\lambda^2\,\hat{\bf z}+w^2\,\br(t)+w\,\bJ(t),
\end{equation}
where $w=u G_1$ is the precession frequency of $\bL_u$ around $\bD_1$. The asymptotic state is $\br=\hat{\bf z}$ and $\bJ=S\, \hat{\bf z}$ and so the conserved length is
$|\bL_u|=|\lambda^2+w^2+w S|$ that we could use to normalize it to unity.

It may seem somewhat surprising that the construction of Eq.~\eqref{eq:Lax-Pendulum} is a vector of conserved length for any~$w$. As an explicit confirmation, we consider $\bL_u^2$, a fourth-order polynomial in $w$, where all five coefficients must be separately conserved. ($w^0$)~The lowest-order term is $\lambda^4$ and thus trivially conserved. ($w^1$)~The linear term is proportional to $J_z$ which is conserved. ($w^2$)~The quadratic term is $2\lambda^2r_z+\bJ^2$ and thus proportional to the conserved total pendulum energy. ($w^3$)~The cubic term is proportional to $S=\bJ\cdot\br$, the conserved pendulum spin. ($w^4$)~The quartic term is $\br^2=1$, completing the proof.

If the starting point is a continuous spectrum $\{v,G_v\}$ with a complex linear eigenfrequency $\Omega$, this information alone is enough to write explicitly the soliton solution for $\bD_1$ and for each individual $\bD_v$ in the form
\begin{equation}\label{eq:Dv-explicit}
    \bD_v(t)=G_v\,\frac{\lambda^2\,\hat{\bf z}+w\,\bJ(t)+w^2\,\br(t)}{\lambda^2+w\,S+w^2},
\end{equation}
where $w=v G_1$. Notice that the fraction is a unit vector that initially points in the positive $z$ direction.

The explicit $x$-$y$-components of this expression are fairly complicated, whereas the $z$ component, encoding the instantaneous amount of flavor conversion, is simply
\begin{equation}
    D^z_v(t)=G_v\,\frac{\lambda^2+w S+w^2\,c(t)}{\lambda^2+w S+w^2},
\end{equation}
where $c(t)$ was provided in Eq.~\eqref{eq:z-2} in terms of the pendulum natural frequency $\lambda$ and spin parameter $\sigma$. The asymptotic value is $c(\pm\infty)=1$ and $D_v(\pm\infty)=G_v$. By construction, the lowest soliton point occurs at $t=0$ and is $c(0)=2\sigma^2-1$. Therefore, the largest excursion of $\bD_v$ from its asymptotic state is
\begin{equation}\label{eq:maximum-excursion}
    D^z_v(0)=G_v\,\frac{\lambda^2+2\lambda\sigma w+(2\sigma^2-1)\,w^2}{\lambda^2+2\lambda\sigma w+w^2},
\end{equation}
where we have used $S=2\lambda\sigma$ and $w=v G_1$.

In Fig.~\ref{fig:example_matter} we have shown the numerical evolution of a certain example, where the spectrum $G_v$ is a thick solid line. The numerical maximum pendulum excursion of the no-matter case, delimiting the shaded region, is equivalent to Eq.~\eqref{eq:maximum-excursion}, as we have explicitly verified. Note that the shown example, Case~D of Ref.~\cite{Padilla-Gay:2021haz}, has $\Omega=1.0743 \pm 1.1121\,i$, which is equivalent to $\lambda=|\Omega|=1.5462$ and spin parameter $\sigma={\rm Re}\,\Omega/|\Omega|=0.6948$.

\subsection{Corollary for Spin Precession}

Spin precession in an external $B$-field is a general topic, for example in Nuclear Magnetic Resonance (NMR) techniques. In neutrino flavor evolution, the propagation through a density profile is equivalent to spin precession in a time varying $B$-field and the MSW effect is the adiabatic version of this effect. Considering the general precession equation $\dot\bP=w\bB\times\bP$, the adiabatic case corresponds to $w$ being large compared with the rate-of-change of the unit vector $\bB(t)$. In this case, the spin follows the $\bB$ field. If it was not initially aligned with $\bB$, this means that its precession cone follows $\bB$ with fixed opening angle.

The results of the previous section imply another special case. If the motion of $\bB(t)$ is equivalent to $\br(t)$ of a soliton defined by $\lambda$ and $\sigma$, then the motion of $\bP$ is once more simple for any value of $w$. Of course, if $w$ is large, Eq.~\eqref{eq:Dv-explicit} reveals explicitly that $\bP$ remains aligned with $\bB$. However, for any $w$, it returns to the asymptotic position, a property of the soliton solution that has nothing directly to do with collective effects. Here we think of $\bB(t)$ as being externally prescribed as in NMR.

Moreover, even if $\bP$ was not initially aligned with $\bB$, its precession cone returns to its original opening angle. So if $\bB$ initially points up and $\bP$ in some arbitrary direction, after $\bB$ returning to its asymptotic upright orientation, the precession cone of $\bP$ has also returned to its initial opening angle. $\bP$ itself has no asymptotic position because it always moves if it is not asymptotically aligned with $\bB$. The opening angle of the precession cone changes during the pendulum motion, in contrast to the adiabatic case, but returns back to its value at early times.

\section{Systems of Reduced Dimensionality}

Whenever we begin with discrete or continuous system $\{v,\bD_v\}$ that is single crossed and provides a single soliton, the true dimensionality of the system is smaller than indicated by its total degrees of freedom. Whenever the linear normal mode analysis produces a complex eigenfrequency $\Omega$, there will be an unstable solution, connecting in the nonlinear regime to a pendulum. We here review several cases of this reduction of dimensionality and arrive at a pendulum that is described by three in dependent Bloch vectors. The aim here is to arrive at a practical way of doing this, which can often be confusing.

\subsection{From Three Beams to Pendulum and Back:
Systematic Lax Vector Approach}

We consider a three beam representation of a bigger system and we imagine them to be Lax vectors of the original system and therefore denote them by $\bL_i$, i.e., the system is $\{u_i,\bL_i\}$ with $i=1,2$, and~3.
The EOMs of the three-beam system are
\begin{equation}
    \dot{\bL}_i=-u_i \bM_1\times\bL_i,
\end{equation}
where we denote the moments by $\bM_n=\sum_{i=1}^3 u_i^n \bL_i$. The moments of the distribution obey the EOMs
\begin{equation}
    \dot{\bM}_n=-\bM_1\times\bM_{n+1}.
\end{equation}
For three-beams, only three of these moments are independent, which we take as $\bM_0$, $\bM_1$, and $\bM_2$. The third moment satisfies
\begin{equation}
    \bM_3= p \bM_0-(u_1 u_2 + u_1 u_3 + u_2 u_3) \bM_1+s \bM_2.
\end{equation}
For brevity, we use $p=u_1u_2u_3$ and $s=u_1+u_2+u_3$. Therefore, the first three moments obey the EOMs
\begin{subequations}
    \begin{eqnarray}
    \dot{\bM}_0&=&0,
    \\
    \dot{\bM}_1&=&-\bM_1\times\bM_2,
    \\
    \dot{\bM}_2&=&-p\bM_1\times\bM_0-s\bM_1\times\bM_2.
\end{eqnarray}
\end{subequations}
The length of the vector $\bM_1$ is conserved, so we may write $\bM_1=|\bM_1| \br$, with $\br$ a unit vector, which will be our pendulum direction. We also introduce the definition $\bJ=\bM_2-s\bM_1$. Finally, we write $\bM_0=|\bM_0| \hat{\bz}$, fixed along the $z$-axis. Therefore, the relevant EOMs become
\begin{equation}
\dot{\br}=\bJ\times\br,
\quad\text{and}\quad
\dot{\bJ}=p|\bM_1||\bM_0|\hat{\bz}\times\br.
\end{equation}
The vector $\bJ$ is the total angular momentum of the pendulum, which we can split into a spin and a orbital part
\begin{equation}
    \bJ=\bL+\bS.
\end{equation}
The orbital angular momentum $\bL$, not to be confused with any of the Lax vectors, is the component of $\bJ$ transverse to $\br$ and is determined by the EOMs~as
\begin{equation}
    \bL=\br\times\dot{\br}.
\end{equation}
On the other hand, the spin $\bS=S\br$ has a length $S$ conserved by the EOMs which can therefore be determined by the initial conditions
\begin{equation}
    S=(\bM_2-s\bM_1)\cdot\br.
\end{equation}
At this point, the only remaining EOM is
\begin{equation}
    \dot{\bL}-S\br\times\bL+p|\bM_1||\bM_0| \br\times\hat{\bz}=0.
\end{equation}
In place of this EOM, one may easily check the existence of two additional integrals of motion, whose conservation completely determines the dynamics; these are the $z$-component of the total angular momentum
\begin{equation}
    J_z=\hat{\bz}\cdot\bL+S\hat{\bz}\cdot\br,
\end{equation}
and the energy
\begin{equation}
    E=\frac{|\bL|^2}{2}+\frac{S^2}{2}+p|\bM_1||\bM_0|\hat{\bz}\cdot\br.
\end{equation}
Comparing with Eqs.~\eqref{eq:pendulumkinetic} and~\eqref{eq:pendulumtotal}, we now find that the dynamics of the three-beam system is indeed identical to the pendulum, with the identification $I=1$ and $Mg\ell=p|\bM_1| |\bM_0|$. In particular, it follows that the single swing of the pendulum introduced in Appendix~\ref{sec:gyropendulum} exactly corresponds to the single soliton in the three-beam system.

To complete the discussion, we finally express the original three beams $\bL_i$ in terms of the pendulum vectors $\bM_0$, $\bM_1=|\bM_1| \br$, and $\bJ$ as
\begin{equation}\label{eq:three_beams_pendulum}
    \bL_i=\frac{\bJ+\bM_1 u_i+\bM_0 \prod_{j\neq i} u_j}{\prod_{j\neq i}(u_i-u_j)}.
\end{equation}
Completing our transformations that mapped our three beams on the pendulum vectors and back.

\subsection{From Continuous Spectrum to Three Beams: 
Systematic Matching Conditions}

As shown in Refs.~\cite{Padilla-Gay:2021haz,Fiorillo:2023mze}, from a three-beam system one can realize a continuous system of polarization vectors with a pendular motion
\begin{equation}\label{eq:full_spectrum_sol}
    \bD_v=\alpha_v \sum_{i=1}^3\frac{u_i\bL_i}{v-u_i},
\end{equation}
where $\alpha_v$ is a set of $v$-dependent constants. This form was motivated in Ref.~\cite{Fiorillo:2023mze}, where we showed that $\bD_v$ evolve as the Lax vectors of the fictitious three-beam systems. As suggested in Ref.~\cite{Padilla-Gay:2021haz}, a practical way of realizing the mapping is: given a continuous spectrum, we identify from linear stability analysis the pendulum parameters, which are determined by the real and imaginary part of the unstable frequency; determine a possible three-beam set which possesses the same pendulum parameters -- or equivalently the same unstable frequency -- and in addition has the same initial value of $\bD_1=\sum_i u_i \bL_i=\sum_v v \bD_v$ and $\bD_0=\sum_i \bL_i=\sum_v \bD_v$; realize the mapping via Eq.~\eqref{eq:full_spectrum_sol}.

A complementary strategy, historically used in the context of slow flavor oscillations~\cite{Raffelt:2011yb}, is to start directly from Eq.~\eqref{eq:full_spectrum_sol}. Since $\bD_v$ are the Lax vectors of the fictitious three-beam systems, they will automatically follow the EOM
\begin{equation}
    \dot{\bD}_v=-v\bM_1\times \bD_v.
\end{equation}
For this subsection, we call $\bM_n=\sum_i u_i^n \bL_i$ the moments of the three-beam system and $\bD_n=\sum_v v^n \bD_v$ the moments of the continuous system. Therefore, in order for the vector $\bD_v$ to obey the correct EOMs, one needs only require the matching of the first moment
\begin{equation}
\bD_1=\bM_1
\end{equation}
at all times.

We notice that, since the dynamics is only regulated by $\bD_1$, it is not strictly necessary to require the matching of the zero moment. Our three-beam system may have a $\bM_0\neq \bD_0$; both $\bM_0$ and $\bD_0$ are conserved by their respective dynamics. In reality, we always have enough freedom to choose our three-beam system such that also the matching $\bM_0=\bD_0$ is realized. However, for clarity, we will first proceed without this requirement, and only later comment on how it can be enforced. We now show that the two strategies -- namely matching of the pendulum parameters, as suggested in Ref.~\cite{Padilla-Gay:2021haz}, and matching of the first moments of the distribution, as suggested in Ref.~\cite{Raffelt:2011yb} -- are completely equivalent.

The matching of the first moment $\bD_1$ requires some algebra, since we need to express the individual vectors $\bL_i$ in terms of the pendulum vectors. Using Eqs.~\eqref{eq:full_spectrum_sol} and~\eqref{eq:three_beams_pendulum}, we write
\begin{equation}\label{eq:expression_full_beams}
    \bD_v=\alpha_v\frac{p\bM_0+v\bJ+v^2 \bM_1}{\prod_i (v-u_i)}.
\end{equation}
Therefore, introducing the family of quantities
\begin{equation}
I_n=\sum_v\frac{\alpha_v v^n}{\prod_i (v-u_i)},
\end{equation}
the matching condition on $\bM_1=\bD_1$ reads
\begin{equation}
p I_1 \bM_0 + I_2 \bJ + I_3 \bM_1=\bM_1.
\end{equation}
Since $\bM_0$ is constant, while the other vectors are not, requiring these conditions to be valid at all times immediately leads to the vanishing of the coefficients of $\bM_0$. Furthermore, since $\bJ$ contains an orbital part $\bL$ which is orthogonal to $\bM_1$, it also follows that its coefficients must separately vanish. Finally, it follows that the coefficients of $\bM_1$ must vanish as well. Therefore, we may write the conditions as
\begin{equation}
    I_1=0,\; I_2=0,\; I_3=1.
\end{equation}
To proceed further, we assume that initially both the three beams and the continuous polarization vectors are closely aligned to the $z$-axis, and we write the constants $\alpha_v$ at the initial time as
\begin{equation}
    \alpha_v=\frac{D^z_v}{\sum_i \frac{L^z_i u_i}{v-u_i}}.
\end{equation}
The integrals $I_n$ therefore become
\begin{equation}
    I_n=\sum_v \frac{D^z_v}{\sum_i\bigl[L^z_i u_i \prod_{j\neq i} (u_j-v)\bigr]}\,v^n.
\end{equation}
The denominator of this expression is a second-order polynomial, which corresponds up to a constant to the initial Lax vector of the three-beam system
\begin{equation}
    \sum_i \Bigl[ L^z_i u_i \prod_{j\neq i} (u_j-v)\Bigr]=\prod_j (v-u_j) \sum_i \frac{L^z_i u_i}{v-u_i}.
\end{equation}
The vanishing of the Lax vector coincides in form with the dispersion relation, see Eq.~\eqref{eq:dispersion_time}. Therefore, it follows that the zeros of the denominator of $I_n$ coincide with the eigenfrequencies of the three-beam system divided by $M^z_1$, namely $\bar{u}=\Omega/M^z_1$. Since the three-beam system is by construction unstable, there are two such frequencies complex conjugate to one another. Therefore, we may rewrite the definition of $I_n$ as
\begin{equation}
    I_n=\sum_v \frac{D^z_v}{M^z_1 (v+\bar{u})(v+\bar{u}^*)} v^n.
\end{equation}
If we now write the two conditions $I_1=0$ and $I_2=0$ as
\begin{subequations}
    \begin{eqnarray}
    \sum_v \frac{v D^z_v}{|v+\bar{u}|^2}&=&0,
    \\
    \sum_v\frac{v D^z_v}{|v+\bar{u}|^2}(v-\mathrm{Re}(u))&=&0,        
    \end{eqnarray}
\end{subequations}
we see that they are equivalent to requiring that the complex eigenfrequency of the continuous beam system coincides with the eigenfrequency of the three-beam system. In turn, the condition $I_3=1$ can be rewritten as
\begin{equation}
    \sum_v \frac{D^z_v v^3}{M^z_1(v^2+|\bar{u}|^2+2v\mathrm{Re}(\bar{u}))}=\sum_v\frac{D^z_v v}{M^z_1}=1,
\end{equation}
and therefore is just the matching of the initial values of $D^z_1$ and $M^z_1$. This confirms that the procedures of Refs.~\cite{Padilla-Gay:2021haz} and~\cite{Raffelt:2011yb} are equivalent.

Therefore, the mapping conditions on $\bD_1$ can be written in the form
\begin{subequations}\label{eq:electrostatic}
\begin{eqnarray}
    &&\sum_v v D^z_v=\sum_i u_i L^z_i,\; \\ 
    \nonumber 
    && \sum_v \frac{v D^z_v\bigl[v-\mathrm{Re}(\bar{u})\bigr]}{v^2+|\bar{u}|^2+2\mathrm{Re}(\bar{u})v}
    \\
    &&\quad{}=\sum_i \frac{u_i L^z_i\,\bigl[u_i-\mathrm{Re}(\bar{u})\bigr]}{u_i^2+|\bar{u}|^2+2\mathrm{Re}(\bar{u})u_i}=0,
    \\ \nonumber 
    && \sum_v \frac{v D^z_v\,\mathrm{Im}(\bar{u})}{v^2+|\bar{u}|^2+2\mathrm{Re}(\bar{u})v}
    \\
    &&\quad{}=\sum_i \frac{u_i L^z_i\,\mathrm{Im}(\bar{u})}{u_i^2+|\bar{u}|^2+2\mathrm{Re}(\bar{u})u_i}=0.
\end{eqnarray}    
\end{subequations}
Incidentally, these conditions can be given a simple physical interpretation using an electrostatic analogy (see also Ref.~\cite{Pehlivan:2011hp}). If we interpret $v D^z_v$ as a continuous linear charge density, the quantity appearing in the second and third Eqs.~\eqref{eq:electrostatic} is the $x$ and $y$ component of the electrostatic field generated at the point $(\mathrm{Re}(\bar{u}),\mathrm{Im}(\bar{u}))$. The charge distribution $v D^z_v$ vanishes at two points, namely at $v=0$ and $v=v_c$, the crossing point of $D^z_v$. Therefore, the three-beam mapping corresponds to replacing the continuous distribution, split into three intervals with alternating charges, by three effective charges $u_i L^z_i$  placed at the positions $u_i$ giving the same total charge, and with an electrostatic field vanishing at the same off-axis point. These effective charges can of course be chosen in the interval $-1<u_i<1$, since they represent some kind of average properties of the continuous distribution in the same interval. 

In addition to these three conditions, we may add the condition on the matching of $\bD_0$ at the initial instant, which reads
\begin{equation}
    \sum_v D^z_v=\sum_i L^z_i.
\end{equation}
In this form, the electrostatic analogy breaks down, since we cannot attribute a direct physical meaning to $D^z_v$.
We notice that the matching condition on $\bD_0$ could also be found by equating the sum over $v$ of Eq.~\eqref{eq:expression_full_beams} to $\bM_0$, which leads to the equality
\begin{equation}
    pI_0=1.
\end{equation}
After multiplying both sides by $|\bar{u}|^2$, and using $I_1=I_2=0$, this equation can be rewritten as
\begin{equation}\label{eq:alternative_condition}
    p\frac{D^z_0}{D^z_1}=|\bar{u}|^2,
\end{equation}
as also found in Ref.~\cite{Padilla-Gay:2021haz}. This is a relation between the unstable frequency, the moments of the distribution, and the three-beam parameters. In particular, it follows that the existence of a three-beam system with all $|u_i|<1$ requires the inequality
\begin{equation}\label{eq:requirement_three_beam}
    |\bar{u}|^2<\left|\frac{D^z_0}{D^z_1}\right|
\end{equation}
to be satisfied. We emphasize that this property is only required if we want to impose the matching both of $\bD_0=\bM_0$ and $\bD_1=\bM_1$; as we have proven earlier, the less restrictive condition $\bD_1=\bM_1$ does not require this property, so that a representative three-beam system can certainly always be found with this relaxed requirement.

Finally, the full set of conditions may be written more compactly as
\begin{equation}
    \sum_v \frac{D^z_v}{|v+\bar{u}|^2} \left(\begin{array}{c} 
  1 \\ v \\ v^2 \\ v^3 \end{array}\right)=\sum_i \frac{L^z_i}{|u_i+\bar{u}|^2} \left(\begin{array}{c} 
  1 \\ u_i \\ u_i^2 \\ u_i^3 \end{array}\right).
\end{equation}
If we regard now $\rho_v=D^z_v/|v+\bar{u}|^2$ as a distribution over the interval $[-1,1]$, this form shows that the three beams are effectively replacing the continuous distribution with a discrete distribution whose moments of order up to $3$ are identical.

\subsection{A practical example}

Here we provide the reader with a worked-out example of how to construct a three-beam system representative of a continuous distribution with a single unstable frequency. Our example is based on the continuous spectrum of case D of Ref.~\cite{Padilla-Gay:2021haz}; the angular distribution is
\begin{equation}
    D^z_v=0.11-0.5 e^{-12.5(1-v)^2}.
\end{equation}
We may first check that our criterion for the existence of an instability, Eq.~\eqref{eq:Nyquist-lab}, is satisfied for this distribution; the crossing is at $v_c=0.652$, and both conditions in Eq.~\eqref{eq:Nyquist-lab} are verified. The unstable frequency that solves the dispersion relation Eq.~\eqref{eq:dispersion_time} is
\begin{equation}
    \bar{u}=\frac{\Omega}{M^z_1}=-0.203994-0.211175 i.
\end{equation}
We can now enforce the conditions Eqs.~\eqref{eq:requirement_three_beam}. These are four equations for the six unknowns $u_i$ and $L_i$. The simplest way to proceed is to notice that the first condition in Eq.~\eqref{eq:requirement_three_beam} can be traded for Eq.~\eqref{eq:alternative_condition}, which provides a definite connection between the velocities of the three beams. As in Ref.~\cite{Padilla-Gay:2021haz}, we may choose $u_1=-1$ and $u_3=1$, so that
\begin{equation}
    u_2=-\frac{D^z_1 |\bar{u}|^2}{D^z_0}.
\end{equation}
For the special Case D as an example, we find
\begin{equation}
    u_2=0.096.
\end{equation}
At this point, our only unknowns are the $L^z_i$, or, equivalently, the combinations 
\begin{equation}
    \Phi_i=\frac{L^z_i}{|u_i+\bar{u}|^2}.
\end{equation}
The three remaining conditions in Eq.~\eqref{eq:requirement_three_beam} form a set of three linear equations in the three unknowns $\Phi_i$, which are therefore easily inverted. Finally, since we know the velocities $u_i$, we can obtain the numerical values for $L^z_i$
\begin{equation}
    L^z_1=0.072,\quad L^z_2=0.062,\quad L^z_3=-0.040.
\end{equation}
Notice that with the choice $u_1=-1$ and $u_3=1$, the procedure leads to a definite prescription for the three-beam system.

\bibliographystyle{bibi}
\bibliography{Biblio}

\end{document}